%% file: main.tex
\documentclass[11pt]{article}
\usepackage{amsmath}
\usepackage{amssymb}
\usepackage{amsfonts}
\usepackage[dvipdfmx]{hyperref}
\usepackage{doi}
\usepackage[backend=bibtex, isbn=false, url=false, style=numeric]{biblatex}
\title{Enumeration of Enumeration Algorithms}
\author{Kunihiro Wasa\footnote{\href{mailto:kunihiro.wasa@gmail.com}{kunihiro.wasa@gmail.com}}}
\begin{document}

\maketitle

\begin{abstract}
In this paper, 
we enumerate enumeration problems and algorithms\footnote{See also \url{http://www-ikn.ist.hokudai.ac.jp/~wasa/enumeration_complexity.html}}. 
Other useful catalogues for enumeration algorithms are provided 
by Komei Fukuda\footnote{\url{http://www-oldurls.inf.ethz.ch/personal/fukudak/publ/Enumeration/enumalgo95_update.pdf}} 
and Yasuko Matsui\footnote{\url{http://www-oldurls.inf.ethz.ch/personal/fukudak/publ/Enumeration/enumalgo93.pdf}}. 
This survey is under construction. 
If you know some results 
not in this survey or there is anything wrong, 
please let me know. 
\end{abstract}

\tableofcontents
\newpage

\input{enum}

\section*{Acknowledgements}
I am deeply grateful to 
Komei Fukuda, 
Leslie Goldberg, 
Christian Komusiewicz, 
Andrea Marino, 
Yasuko Matsui, 
Valia Mitsou, 
Shin-ichi Nakano, 
Ryuhei Uehara, 
and Katsuhisa Yamanaka for enumerating.

\printbibliography
\end{document}

%% file: enum.tex
\section{Geometry}
     \subsection{Arrangement}
        \subsubsection{Enumeration of all cells in arrangements}
            \begin{description}
                \item[Input] An arrangement of distinct hyperplanes.
                \item[Output] All cells in arrangements.
                \item[Complexity] $O(nm\ell(n,m)N)$ total time and $O(nm)$ space.
                \item[Comment] $n$ is the dimension, $m$ is the number of hyperplanes, and $\ell(n, m)$ is the time for solving an LP with $n$ variables and $m-1$ inequalities.  $N$ is the number of solutions.
                \item[Reference] \cite{Avis1996}
            \end{description}
     \subsection{Face}
        \subsubsection{Enumeration of all arrangements}
            \begin{description}
                \item[Input] An integer $n$ and $\mathcal{J} \subseteq \{+, -\}^n$.
                \item[Output] The set $\mathcal{F}$ of faces if $\mathcal{J}$ is the set of maximal faces of an oriented matroid
                \item[Complexity] $O(\max(n^2 |\mathcal{J}|^3, n^3 |\mathcal{J}|^2))$ total time.
                \item[Comment] Their paper treats a reconstruction of the location vectors of all faces from the graph.
                \item[Reference] \cite{Fukuda1991}
            \end{description}
     \subsection{Matching}
        \subsubsection{Enumeration of all non-crossing perfect matchings (and convex partitions) in a given points}
            \begin{description}
                \item[Input] A point sets $P$ in a plane
                \item[Output] All non-crossing perfect matchings (and convex partitions) in $P$. 
                \item[Complexity] After $O(2^nn^4)$ preprocessing time, polynomial delay. 
                \item[Comment] $n$ is the number of points in $P$
                \item[Reference] \cite{Wettstein2014}
            \end{description}
     \subsection{Nearest neighbor}
        \subsubsection{Enumeration of the $k$ smallest distances between pairs of points}
            \begin{description}
                \item[Input] $n$ points in a plane.
                \item[Output] The $k$ smallest distances between pairs of points in non-decreasing order.
                \item[Complexity] $O(n \log n + k \log k)$ total time and $O(n + k)$ space.
                \item[Reference] \cite{Dickerson1992}
            \end{description}
     \subsection{Spanning tree}
        \subsubsection{Enumeration of the Euclidean $k$ best spanning trees in a plane with $n$ points}
            \begin{description}
                \item[Input] $n$ points in a plane
                \item[Output] The Euclidean $k$ best spanning trees of the given points.
                \item[Complexity] $O(k^2 n + n \log n)$ total time
                \item[Comment] An Euclidean spanning tree in a plane is a spanning tree of the complete graph whose vertices are all given points and the weight of an edge equal to the Euclidean distance between the corresponding vertices.
                \item[Reference] \cite{Eppstein1992}
            \end{description}
        \subsubsection{Enumeration of the orthogonal $k$ best spanning trees in a plane with $n$ points}
            \begin{description}
                \item[Input] $n$ points in a plane
                \item[Output] The orthogonal $k$ best spanning trees of the given points.
                \item[Complexity] $O(k^2 + k n \log \log (n/k) + n \log n)$ total time.
                \item[Comment] An orthogonal spanning tree in a plane is a spanning tree of the complete graph whose vertices are all given points and the weight of an edge equal to the $L_1$ distance between the corresponding vertices.
                \item[Reference] \cite{Eppstein1992}
            \end{description}
        \subsubsection{Enumeration of all spanning trees of a plane}
            \begin{description}
                \item[Input] A point set $P$.
                \item[Output] All spanning trees of $P$.
                \item[Complexity] $O(|P|^3N)$ total time and $O(|P|)$ space.
                \item[Comment] $N$ is the number of solutions.
                \item[Reference] \cite{Avis1996}
            \end{description}
        \subsubsection{Enumeration of the Euclidean $k$ best spanning trees in a plane with $n$ points}
            \begin{description}
                \item[Input] $n$ points in a plane
                \item[Output] The Euclidean $k$ best spanning trees of the given points.
                \item[Complexity] $O(n \log n \log k + k \min(k, n)^{1/2})$ total time
                \item[Comment] An Euclidean spanning tree in a plane is a spanning tree of the complete graph whose vertices are all given points and the weight of an edge equal to the Euclidean distance between the corresponding vertices.
                \item[Reference] \cite{Eppstein1997}
            \end{description}
     \subsection{Triangulation}
        \subsubsection{Enumeration of all regular triangulations}
            \begin{description}
                \item[Input] $n$ points.
                \item[Output] All regular triangulations of given points in $(d-1)$-dimensional space.
                \item[Complexity] $O(dsLP(r, ds)T)$ time and $O(ds)$ space.
                \item[Comment] $s$ is the upper bound of the number of simplcies contained in one regular triangulation, $LP(r, ds)$ denotes the time complexity of solving linear programming problem with $ds$ strict inequality constraints in $r$ variables, and $T$ is the number of regular triangulations.
                \item[Reference] \cite{Masada1996}
            \end{description}
        \subsubsection{Enumeration of all spanning regular triangulations}
            \begin{description}
                \item[Input] $n$ points.
                \item[Output] All spanning regular triangulations of given points in $(d-1)$-dimensional space.
                \item[Complexity] $O(dsLP(r, ds)T')$ time and $O(ds)$ space.
                \item[Comment] $s$ is the upper bound of the number of simplcies contained in one regular triangulation, $LP(r, ds)$ denotes the time complexity of solving linear programming problem with $ds$ strict inequality constraints in $r$ variables, and $T'$ is the number of regular triangulations.
                \item[Reference] \cite{Masada1996}
            \end{description}
        \subsubsection{Enumeration of all triangulations of a point set}
            \begin{description}
                \item[Input] A point set $P$.
                \item[Output] All triangulations of $P$.
                \item[Complexity] $O(|P|N)$ total time and $O(|P|)$ space.
                \item[Comment] $N$ is the number of solutions.
                \item[Reference] \cite{Avis1996}
            \end{description}
        \subsubsection{Enumeration of all triangulations in general dimensions}
            \begin{description}
                \item[Input] Points.
                \item[Output] All triangulations.
                \item[Complexity] See the paper.
                \item[Comment] This algorithm uses the enumeration algorithm for maximal independent sets.
                \item[Reference] \cite{Takeuchi}
            \end{description}
        \subsubsection{Enumeration of all regular triangulations in general dimensions}
            \begin{description}
                \item[Input] Points.
                \item[Output] All regular triangulations.
                \item[Complexity] See the paper.
                \item[Comment] This algorithm uses the enumeration algorithm for maximal independent sets.
                \item[Reference] \cite{Takeuchi}
            \end{description}
        \subsubsection{Enumeration of all triangulation paths of a point set}
            \begin{description}
                \item[Input] A point set $P$.
                \item[Output] All triangulation paths of $P$.
                \item[Complexity] $O(N|P|^3\log |P|)$ time and $O(|P|)$ space.
                \item[Comment] $N$ is the number of solutions.
                \item[Reference] \cite{Dumitrescu2001}
            \end{description}
        \subsubsection{Enumeration of all pseudotriangulations of a finite point set}
            \begin{description}
                \item[Input] A point set $S$ of size $n$.
                \item[Output] All pointed pseudotriangulations of $S$.
                \item[Complexity] $O(\log n)$ time per solution with linear space.
                \item[Reference] \cite{Bereg2005}
            \end{description}
        \subsubsection{Enumeration of all pseudotriangulations of a point set}
            \begin{description}
                \item[Input] A point set $P$.
                \item[Output] All pseudotriangulations of $P$.
                \item[Complexity] $O(n)$ time per pseudotriangulation.
                \item[Comment] $n$ is the number of points in $P$.
                \item[Reference] \cite{Bronnimann2006a}
            \end{description}
        \subsubsection{Enumeration of all triangulations of a convex polygon of $n$ vertices with numbered}
            \begin{description}
                \item[Input] A convex polygon $P$ with $n$ vertices that are numbered.
                \item[Output] All triangulations of $P$.
                \item[Complexity] $O(1)$ time per triangulation and $O(n)$ space.
                \item[Reference] \cite{Parvez2009}
            \end{description}
        \subsubsection{Enumeration of all triangulations of a convex polygon of $n$ vertices}
            \begin{description}
                \item[Input] A convex polygon $P$ with $n$ vertices that are not numbered.
                \item[Output] All triangulations of $P$.
                \item[Complexity] $O(n^2)$ time per triangulation and $O(n)$ space.
                \item[Reference] \cite{Parvez2009}
            \end{description}
\section{Graph}
     \subsection{$k$-degenerate graph}
        \subsubsection{Enumerate well-ordered (strongly) $k$-generate with $n$ vertices}
            \begin{description}
                \item[Input] $n$: the number of vertices.
                \item[Output] All well-ordered (strongly) $k$-degenerate graphs with $n$ vertices.
                \item[Complexity] $O(nm+m^2)$ time per enumerated and printed graph.
                \item[Comment] $m$ is the number of edges of printed graphs.
                \item[Reference] \cite{Bauer2010a}
            \end{description}
        \subsubsection{Enumerate well-ordered (strongly) $k$-generate with $n$ vertices and $m$ edges}
            \begin{description}
                \item[Input] $n$: the number of vertices, $m$: the number of edges.
                \item[Output] All well-ordered (strongly) $k$-degenerate graphs with $n$ vertices and $m$ edges.
                \item[Complexity] $O(n^{3/2}m^2)$ time per enumerated and printed graph.
                \item[Reference] \cite{Bauer2010a}
            \end{description}
     \subsection{Bounded union}
        \subsubsection{Enumeration of all bounded union}
            \begin{description}
                \item[Input] A graph with $h$ multiedges each with at most $d$ vertices.
                \item[Output] All minimal subset $U$ of at most $k$ vertices that entirely includes at least one set from each multiedge.
                \item[Complexity] $O(dc^{k+1}h + \min(kc^{2k}, hkc^k))$ time.
                \item[Reference] \cite{Damaschke2006}
            \end{description}
     \subsection{Bridge}
        \subsubsection{Enumeration of all bridge-avoiding extensions of a graph}
            \begin{description}
                \item[Input] A graph $G = (V, E)$ and an edge subset $B \subseteq E$.
                \item[Output] All bridge-avoiding extensions of $G$.
                \item[Complexity] $O(K^2 \log(K) |E|^2 + K^2(|V|+|E|)|E|^2)$ total time.
                \item[Comment] $K$ is the number of bridge-avoiding extensions.  $X$ is a bridge-avoiding extensions of $G$ if $X$ is a minimal edge set $X \subseteq E \setminus B$ such that no edge $b \in B$ is a bridge in $(V, B\cup X)$.
                \item[Reference] \cite{Khachiyan2008a}
            \end{description}
     \subsection{Bubble}
        \subsubsection{Enumeration of all bubbles in a directed graph}
            \begin{description}
                \item[Input] A directed graph $G=(V, E)$ and a source $s$.
                \item[Output] All bubbles with a given source $s$.
                \item[Complexity] $O(|V| + |E|)$ delay with $O(|V|(|V| + |E|))$ preprocessing time.
                \item[Comment] An $(s,t)$-bubble is two disjoint $(s,t)$-paths that only share $s$ and $t$.
                \item[Reference] \cite{Birmele2012}
            \end{description}
        \subsubsection{Enumeration of all $(s, t, \alpha_1, \alpha_2)$-bubble in a directed graph}
            \begin{description}
                \item[Input] A directed graph $G = (V, E)$, source vertex $s$, and two upper bounds $\alpha_1$ and $\alpha_2$.
                \item[Output] All $(s, t, \alpha_1, \alpha_2)$-bubble in $G$.
                \item[Complexity] $O(|V|(|E|+|V|\log |V|))$ delay.
                \item[Comment] A pair of two vertex-disjoint $(s, t)$-paths $p_1$ and $p_2$ is $(s, t, \alpha_1, \alpha_2)$-bubble in $G$, if $|p_1| \le \alpha_1$ and $|p_2| \le \alpha_2$.
                \item[Reference] \cite{Sacomoto2013b}
            \end{description}
     \subsection{Chord diagram}
        \subsubsection{Enumeration of all non-isomorphic chord diagrams}
            \begin{description}
                \item[Input] An integer $2n$.
                \item[Output] All non-isomorphic chord diagrams with $2n$ points.
                \item[Comment] A \textit{chord diagram} is a set of $2n$ points on an oriented circle (counterclockwise) joined pairwise by $n$ chords.  In an experiment, their algorithm runs in almost CAT, but there is no Mathematical proof.
                \item[Reference] \cite{Sawada2002a}
            \end{description}
     \subsection{Chordal graph}
        \subsubsection{Enumeration of all chordal graphs in a graph}
            \begin{description}
                \item[Input] A graph $G$.
                \item[Output] All chordal graphs in $G$.
                \item[Complexity] $O(1)$ delay with $O(n^3)$ space.
                \item[Comment] Using reverse search.
                \item[Reference] \cite{Kiyomi2006}
            \end{description}
        \subsubsection{Enumeration of all chordal supergraph that contains a given graph}
            \begin{description}
                \item[Input] A graph $G = (V, E)$.
                \item[Output] All chordal supergraphs each of which contain $G$.
                \item[Complexity] $O(|V|^3)$ time for each and $O(|V|^2)$ space.
                \item[Reference] \cite{Kiyomi2006a}
            \end{description}
        \subsubsection{Enumeration of all chordal sandwiches in a graph}
            \begin{description}
                \item[Input] A graph $G = (V, E)$.
                \item[Output] All chordal sandwiches in $G$.
                \item[Complexity] Polynomial delay with polynomial space.
                \item[Reference] \cite{Kijima2010}
            \end{description}
     \subsection{Clique}
        \subsubsection{Enumeration of all cliques in a graph}
            \begin{description}
                \item[Input] A graph $G$. 
                \item[Output] All cliques in $G$. 
                \item[Comment] They pointed out Augustson and Minker's algorithm has two errors. 
                \item[Reference] \cite{Mulligan1972}
            \end{description}
        \subsubsection{Enumeration of all maximal clique in a graph}
            \begin{description}
                \item[Input] An undirected graph $G=(V, E)$.
                \item[Output] All maximal clique.
                \item[Reference] \cite{Akkoyunlu1973}
            \end{description}
        \subsubsection{Enumeration of all cliques in a graph}
            \begin{description}
                \item[Input] A graph $G$.
                \item[Output] All cliques in $G$.
                \item[Reference] \cite{Bron1973}
            \end{description}
        \subsubsection{Enumeration of all cliques in an undirected graph}
            \begin{description}
                \item[Input] An undirected graph $G$. 
                \item[Output] All cliques in $G$. 
                \item[Reference] \cite{Johnston1976}
            \end{description}
        \subsubsection{Enumeration of all maximal cliques in a given graph}
            \begin{description}
                \item[Input] A graph $G$.  
                \item[Output] All maximal cliques in $G$.
                \item[Reference] \cite{Gerhards1979}
            \end{description}
        \subsubsection{Enumeration of all maximum cliques in a circle graph}
            \begin{description}
                \item[Input] A circle graph $G = (V, E)$.
                \item[Output] All maximum cliques in $G$.
                \item[Complexity] $O(1)$ time per maximum clique.
                \item[Comment] A graph is a circle graph if it is the intersection graph of chords in a circle. The definition of a circle graph can be found in Information System on Graph Classes and their Inclusions\footnote{\url{http://www.graphclasses.org/classes/gc\_132.html}}.
                \item[Reference] \cite{Rotem1981}
            \end{description}
        \subsubsection{Enumeration of all cliques in a connected graph}
            \begin{description}
                \item[Input] A connected graph $G = (V, E)$ and an order $l \ge 2$ of cliques.
                \item[Output] All cliques in $G$.
                \item[Complexity] $O(l\alpha(G)^{l-2}|E|)$ total time and linear space.
                \item[Comment] $\alpha(G)$ is the minimum number of edge-disjoint spanning forests into which $G$ can be decomposed.
                \item[Reference] \cite{Chiba1985}
            \end{description}
        \subsubsection{Enumeration of all maximal cliques in a connected graph}
            \begin{description}
                \item[Input] A connected graph $G = (V, E)$.
                \item[Output] All maximal cliques in $G$.
                \item[Complexity] $O(\alpha(G)|E|)$ total time and linear space.
                \item[Comment] $\alpha(G)$ is the minimum number of edge-disjoint spanning forests into which $G$ can be decomposed.
                \item[Reference] \cite{Chiba1985}
            \end{description}
        \subsubsection{Enumeration of all maximum cliques of a circular-arc graph}
            \begin{description}
                \item[Input] A circular-arc graph $G =(V, E)$.
                \item[Output] All maximum cliques of $G$.
                \item[Complexity] $O(|V|^{3.5} + N)$.
                \item[Comment] $N$ is the number of maximum cliques of $G$.  For a family $A$ of arcs on a circle, a graph $G = (V, E)$ is called the \textit{circular-arc graph} for $A$ if there exists a one-to-one correspondence between $V$ and $A$ such that two vertices in $V$ are adjacent if and only if their corresponding arcs in $A$ intersect.
                \item[Reference] \cite{Kashiwabara1992}
            \end{description}
        \subsubsection{Enumeration of all maximal cliques in a graph}
            \begin{description}
                \item[Input] A graph $G=(V,E)$.
                \item[Output] All maximal cliques included in $G$.
                \item[Complexity] $O(M(n))$ time delay and $O(n^2)$ space, or $O(\delta^4)$ time delay and $O(n+m)$ space, after $O(nm)$ preprocessing.
                \item[Comment] $M(n)$: the time needed to multiply two $n \times n$ matrices, $\delta$: maximum degree of $G=(V,E)$, $n$: total number of vertices, and $m$: total number of edges.
                \item[Reference] \cite{Makino2004}
            \end{description}
        \subsubsection{Enumeration of all maximal cliques in a bipartite graph}
            \begin{description}
                \item[Input] A bipartite graph $G=(V,E)$.
                \item[Output] All maximal bicliques included in $G$.
                \item[Complexity] $O(M(n))$ time delay and $O(n^2)$ space, or $O(\delta^4)$ time delay and $O(n+m)$ space, after $O(nm)$ preprocessing.
                \item[Comment] $M(n)$: the time needed to multiply two $n \times n$ matrices, $\delta$: maximum degree of $G=(V,E)$, $n$: total number of vertices, and $m$: total number of edges.
                \item[Reference] \cite{Makino2004}
            \end{description}
        \subsubsection{Enumeration of all maximal cliques in a dynamic graph}
            \begin{description}
                \item[Input] A graph $G_t = (V, E_t)$.
                \item[Output] All maximal cliques in $G_t$.
                \item[Reference] \cite{Stix2004}
            \end{description}
        \subsubsection{Enumeration of all maximal cliques in a dynamic graph}
            \begin{description}
                \item[Input] A dynamic graph $G$.
                \item[Output] All maximal cliques in $G$.
                \item[Reference] \cite{Stix2004}
            \end{description}
        \subsubsection{Enumeration of all bicliques in a graph in lexicographical order}
            \begin{description}
                \item[Input] A graph $G = (V, E)$.
                \item[Output] All bicliques in $G$.
                \item[Complexity] $O(|V|^3)$ delay and $O(2^{|V|})$ space.
                \item[Comment] There is no polynomial-delay enumeration algorithm for all bicliques in reverse lexicographical order unless $P = NP$.
                \item[Reference] \cite{Dias2005}
            \end{description}
        \subsubsection{Enumeration of all maximal cliques in a graph}
            \begin{description}
                \item[Input] A graph $G = (V, E )$.
                \item[Output] All maximal Cliques in $G$.
                \item[Complexity] $O(\frac{\Delta M_c Tri^2}{P})$ delay and $O(|V| + |E|)$ space.
                \item[Comment] $\Delta$ is the maximum degree in $G$, $M_c$ is the size of the maximum clique in $G$, and $Tri$ is the number of triangles in $G$. This algorithm runs on the computer with $P$ processors.
                \item[Reference] \cite{Du2009}
            \end{description}
        \subsubsection{Enumeration of all cliques in a chordal graph}
            \begin{description}
                \item[Input] A chordal graph $G$.
                \item[Output] All cliques in $G$.
                \item[Complexity] $O(1)$ time per solution on average.
                \item[Reference] \cite{Kiyomi2006}
            \end{description}
        \subsubsection{Enumeration of all maximal cliques in a graph}
            \begin{description}
                \item[Input] A graph $G = (V, E)$.
                \item[Output] All maximal cliques in $G$.
                \item[Complexity] $O(3^{n/3})$ total time.
                \item[Reference] \cite{Tomita2006}
            \end{description}
        \subsubsection{Enumeration of $K$-largest maximal cliques in a graph}
            \begin{description}
                \item[Input] A graph $G$.
                \item[Output] $K$-largest maximal cliques in $G$.
                \item[Reference] \cite{Bulo2007}
            \end{description}
        \subsubsection{Enumeration of all maximal cliques in a graph}
            \begin{description}
                \item[Input] A graph $G$.
                \item[Output] All maximal cliques in $G$.
                \item[Reference] \cite{Pan2008}
            \end{description}
        \subsubsection{Enumeration of all maximal bicliques in a bipartite graph}
            \begin{description}
                \item[Input] A bipartite graph $G = (U, V, E)$.
                \item[Output] All maximal bicliques in $G$ in lexicographical order on $U$.
                \item[Complexity] $O((|U| + |V|)^2)$ delay with exponential space.
                \item[Reference] \cite{Gely2009}
            \end{description}
        \subsubsection{Enumeration of all maximal cliques in a comparability graph}
            \begin{description}
                \item[Input] A comparability graph $G = (V, E)$.
                \item[Output] All maximal cliques in $G$ in lexicographical order.
                \item[Complexity] $O(|V|)$ delay and $O(|V| + |E|)$ space.
                \item[Reference] \cite{Gely2009}
            \end{description}
        \subsubsection{Enumeration of all maximal cliques in a graph}
            \begin{description}
                \item[Input] A graph $G = (V, E)$.
                \item[Output] All maximal cliques in $G$ in lexicographical order.
                \item[Complexity] $O(|V||E|)$ delay with exponential space.
                \item[Reference] \cite{Gely2009}
            \end{description}
        \subsubsection{Enumeration of all maximal cliques in a graph}
            \begin{description}
                \item[Input] A graph $G$.
                \item[Output] All maximal cliques in $G$.
                \item[Comment] This algorithm can be paralleled.
                \item[Reference] \cite{Schmidt2009}
            \end{description}
        \subsubsection{Enumeration of all $c$-isolated maximal clique in a graph}
            \begin{description}
                \item[Input] A graph $G = (V, E)$ and a positive real number $c$.
                \item[Output] All $c$-isolated maximal clique in $G$.
                \item[Complexity] $O(c^42^{2c}|E|)$ total time.
                \item[Reference] \cite{Ito2009}
            \end{description}
        \subsubsection{Enumeration of all $c$-isolated pseudo-clique in a graph}
            \begin{description}
                \item[Input] A graph $G = (V, E)$ and a positive real number $c$.
                \item[Output] All $c$-isolated $PC(k-\log k, \frac{k}{\log k})$ in $G$.
                \item[Complexity] $O(c^42^{2c}|E|)$ total time.
                \item[Comment] $PC(\alpha, \beta)$ is a induced subgraph of $G$ with an average degree at least $\alpha$ and the minimum degree at least $\beta$.
                \item[Reference] \cite{Ito2009}
            \end{description}
        \subsubsection{Enumeration of all $c$-isolated maximal cliques in a graph}
            \begin{description}
                \item[Input] A graph $G = (V, E)$ and an integer $c$.
                \item[Output] All avg-$c$-isolated maximal cliques in $G$.
                \item[Complexity] $O(2^cc^5|E|)$ total time.
                \item[Comment] A vertex set $S \subseteq V$ with $k$ vertices is $c$-isolated if it has less than $ck$ outgoing edges, where an outgoing edge is an edge between a vertex in $S$ and a vertex in $V \setminus S$.
                \item[Reference] \cite{Huffner2009}
            \end{description}
        \subsubsection{Enumeration of all max-$c$-isolated maximal cliques in a graph}
            \begin{description}
                \item[Input] A graph $G = (V, E)$ and an integer $c$.
                \item[Output] All max-$c$-isolated maximal cliques in $G$.
                \item[Complexity] $O(2^cc^5|E|)$ total time.
                \item[Comment] A vertex set $S \subseteq V$ with $k$ vertices is max-$c$-isolated if every vertex in $S$ has less than $c$ neighbors in $V \setminus S$.
                \item[Reference] \cite{Huffner2009}
            \end{description}
        \subsubsection{Enumeration of all maximal $c$-isolated cliques in a graph}
            \begin{description}
                \item[Input] A graph $G = (V, E)$ and an integer $c$.
                \item[Output] All maximal $c$-isolated cliques in $G$.
                \item[Complexity] $O(2.89^cc^2|E|)$ total time.
                \item[Comment] A vertex set $S \subseteq V$ with $k$ vertices is $c$-isolated if it has less than $ck$ outgoing edges, where an outgoing edge is an edge between a vertex in $S$ and a vertex in $V \setminus S$.
                \item[Reference] \cite{Komusiewicz2009}
            \end{description}
        \subsubsection{Enumeration of all cliques in a graph with degeneracy $d$}
            \begin{description}
                \item[Input] A graph $G = (V, E)$ with degeneracy $d$.
                \item[Output] All cliques in $G$.
                \item[Complexity] $O(d|V|3^{d/3})$ time.
                \item[Comment] They also show the largest possible number of maximal cliques in $G$. The number is $(|V|-d)3^{d/3}$.
                \item[Reference] \cite{Eppstein2010}
            \end{description}
        \subsubsection{Enumeration of all pseudo-cliques in a graph}
            \begin{description}
                \item[Input] A graph $G = (V, E)$ and a threshold $\theta$.
                \item[Output] All pseudo-cliques in $G$, whose densities are no less than $\theta$.
                \item[Complexity] $O(\Delta |V| + \mathrm{min}\{\Delta^2, |V| + |E|\})$ delay with $O(|V| + |E|)$ space.
                \item[Comment] $\Delta$ is the maximum degree of $G$. The density of a subgraph $G[S]$ induced by $S \subseteq V$ is given by the number of edges over the number of all its vertex pairs.
                \item[Reference] \cite{Uno2008a}
            \end{description}
        \subsubsection{Enumeration of all maximal cliques in a graph with limited memory}
            \begin{description}
                \item[Input] A graph $G$.
                \item[Output] All maximal graphs in $G$.
                \item[Complexity] See the paper.
                \item[Reference] \cite{Cheng2012b}
            \end{description}
        \subsubsection{Enumeration of all maximal cliques in a graph}
            \begin{description}
                \item[Input] An undirected graph $G=(V, E)$.
                \item[Output] All maximal clique.
                \item[Complexity] $O(\delta\cdot H^3)$ time delay and $O(n+m)$ space.
                \item[Comment] $H$ satisfies $|\{v\in V | \sigma(v)\ge H\}| \le H$.
                \item[Reference] \cite{Chang2013}
            \end{description}
        \subsubsection{Enumeration of all maximal cliques in a graph}
            \begin{description}
                \item[Input] A graph $G$.
                \item[Output] All maximal cliques in $G$.
                \item[Reference] \cite{Henry2013}
            \end{description}
     \subsection{Coloring}
        \subsubsection{Enumeration of all the edge colorings in a bipartite graph}
            \begin{description}
                \item[Input] A bipartite graph $G = (V, E)$.
                \item[Output] All edge colorings in $G$.
                \item[Complexity] $O(\Delta|E|)$ time per solution and space.
                \item[Comment] $\Delta$ is the maximum degree in $G$.
                \item[Reference] \cite{Yasuko1994}
            \end{description}
        \subsubsection{Enumeration of all the edge colorings of a bipartite graph}
            \begin{description}
                \item[Input] A bipartite graph $B = (V, E)$.
                \item[Output] All the edge colorings pf $B$.
                \item[Complexity] $O(T(|V|+|E|+\Delta) + K \min \{|V|^2 + |E|, T(|V|, |E|, \Delta)\})$ total time and $O(|E|\Delta)$ space.
                \item[Comment] $\Delta$ is the number of maximum degree and $T(|V|, |E|, \Delta)$ is the time complexity of an edge coloring algorithm.
                \item[Reference] \cite{Matsui1996b}
            \end{description}
        \subsubsection{Enumeration of all the edge colorings of a bipartite graph}
            \begin{description}
                \item[Input] A bipartite graph $B = (V, E)$.
                \item[Output] All the edge colorings pf $B$.
                \item[Complexity] $O(|V|)$ time per solution and $O(|E|)$ space.
                \item[Reference] \cite{Matsui1996a}
            \end{description}
     \subsection{Connected}
        \subsubsection{Enumeration of all minimal strongly connected subgraphs in a strongly connected subgraph}
            \begin{description}
                \item[Input] A strongly connected subgraph $G$.
                \item[Output] All minimal strongly connected subgraph.
                \item[Complexity] Incremental polynomial time.
                \item[Reference] \cite{Boros2004b}
            \end{description}
        \subsubsection{Enumeration of all minimal 2-vertex connected spanning subgraphs in a graph}
            \begin{description}
                \item[Input] A graph $G$.
                \item[Output] All minimal 2-vertex connected spanning subgraphs of $G$.
                \item[Complexity] Incremental polynomial time.
                \item[Reference] \cite{Khachiyan2006a}
            \end{description}
     \subsection{Cut}
        \subsubsection{Enumeration of all cuts between all pair of vertices in a given graph}
            \begin{description}
                \item[Input] A graph $G$. 
                \item[Output] All cuts bewteen all pair of vertices in $G$
                \item[Reference] \cite{Martelli1976a}
            \end{description}
        \subsubsection{Enumeration of all cutsets in a graph}
            \begin{description}
                \item[Input] A graph $G = (V, E)$.
                \item[Output] All cutsets in $G$.
                \item[Complexity] $O((|V| + |E|)(\mu + 1))$ total time and $O(|V| + |E|)$ or $O(|V|^2)$ space.
                \item[Comment] $\mu$ is the number of solutions.
                \item[Reference] \cite{Tsukiyama1980}
            \end{description}
        \subsubsection{Enumeration of $k$ best cuts in a directed graph}
            \begin{description}
                \item[Input] A directed graph $G = (V, E)$.
                \item[Output] $k$ best cuts in $G$.
                \item[Complexity] $O(k|V|^4)$ total time.
                \item[Reference] \cite{Hamacher1984a}
            \end{description}
        \subsubsection{Enumeration of $K$-best cuts in a network}
            \begin{description}
                \item[Input] A graph $G = (V, E)$.
                \item[Output] $K$-best cuts in $G$.
                \item[Complexity] $O(K\cdot |V|^4)$ time.
                \item[Reference] \cite{Hamacher1984a}
            \end{description}
        \subsubsection{Enumeration of all articulation pairs in a planar graph}
            \begin{description}
                \item[Input] An undirected graph $G = (V, E)$.
                \item[Output] All articulation pairs in $G$.
                \item[Complexity] $O(|V|^2)$ total time.
                \item[Reference] \cite{Laumond1985}
            \end{description}
        \subsubsection{Enumeration of all minimum-size separating vertex sets in a graph}
            \begin{description}
                \item[Input] A graph $G = (V, E)$.
                \item[Output] All minimum-size separating vertex sets.
                \item[Complexity] $\Theta(M|V| + C) = O(2^kn^3)$ total time.
                \item[Comment] $M$ is the number of solutions, $k$ is the connectivity of $G$, and $C = k|V| \min(k(|V|+|E|), A)$, where $A$ is the time complexity of the best maximum network flow algorithm for unit network.
                \item[Reference] \cite{Kanevsky1993}
            \end{description}
        \subsubsection{Enumeration of all minimal separators in a graph}
            \begin{description}
                \item[Input] A graph $G = (V, E)$.
                \item[Output] All minimal separators in $G$.
                \item[Complexity] $O(|V|^6R)$ total time.
                \item[Comment] $R$ is the number of solutions.
                \item[Reference] \cite{Kloks1994}
            \end{description}
        \subsubsection{Enumeration of all $(s, t)$-cuts in a graph}
            \begin{description}
                \item[Input] A graph $G = (V, E)$ and two vertices $s, t$ in $G$.
                \item[Output] All $(s, t)$-cuts in $G$.
                \item[Complexity] $O(|E|)$ time per cut.
                \item[Reference] \cite{Provan1996}
            \end{description}
        \subsubsection{Enumeration of all $(s, t)$-cuts in a directed graph}
            \begin{description}
                \item[Input] A directed graph $G = (V, E)$ and two vertices $s, t$ in $G$.
                \item[Output] All $(s, t)$-cuts in $G$.
                \item[Complexity] $O(|E|)$ time per cut.
                \item[Reference] \cite{Provan1996}
            \end{description}
        \subsubsection{Enumeration of all $(s, t)$-uniformly directed cuts in a directed graph}
            \begin{description}
                \item[Input] A directed graph $G = (V, E)$ and two vertices $s, t$ in $G$.
                \item[Output] All $(s, t)$-uniformly directed cuts in $G$.
                \item[Complexity] $O(|V|)$ time per cut.
                \item[Comment] An \textit{undirected directed cut} is also called a UDC. An $(s, t)$-DUC is an $(s, t)$-cut $(X, Y)$ such that $(X, Y) = \emptyset$, where $(X, Y) = \{(u, v) \in E: u\in X, v\in Y\}$.
                \item[Reference] \cite{Provan1996}
            \end{description}
        \subsubsection{Enumeration of all minimum weighted $(s, t)$ cuts in an weighted graph}
            \begin{description}
                \item[Input] An weighted graph $G = (V, E)$ and two vertices $s, t$ in $G$.
                \item[Output] All minimum weighted $(s, t)$ cuts in $G$.
                \item[Complexity] $O(|V|)$ time per cut.
                \item[Reference] \cite{Provan1996}
            \end{description}
        \subsubsection{Enumeration of all semidirected $(s, t)$ cuts in a directed graph}
            \begin{description}
                \item[Input] A directed graph $G = (V, E)$ and two vertices $s, t$ in $G$.
                \item[Output] All semidirected $(s, t)$ cuts in $G$.
                \item[Complexity] $O(|V||E|)$ time per cut.
                \item[Comment] For a subset $D$ of directed edges, a \textit{semidirected $(s, t)$ cut} with respect to $D$ is an $(s, t)$ cut $(X, Y)$ such that $(X, Y) \cup (Y, X)$ defnes an undirected $(s, t)$ cutset and such that $(X, Y) \cap D = \emptyset$, where a \textit{cutset} is an minimal cut set.
                \item[Reference] \cite{Provan1996}
            \end{description}
        \subsubsection{Enumeration of all strong $(s, K)$ cutsets in a graph}
            \begin{description}
                \item[Input] A graph $G = (V, E)$ , $s\in V$ and $K \subseteq V \setminus \{s\}$.
                \item[Output] All strong $(s, K)$ cutsets in $G$.
                \item[Complexity] $O(|E|)$ time per cut.
                \item[Comment] An \textit{$(s, K)$-cut} is defined to be any cut $(X, Y)$ for which $s \in X$ and $K \cap Y \neq \emptyset$. A \textit{strong $(s, K)$ cutset} is minimal cuts of the form $(X, Y)$ where $s \in X$ and $K \subseteq Y$.
                \item[Reference] \cite{Provan1996}
            \end{description}
        \subsubsection{Enumeration of all strong $(s, K)$ cutsets in a directed graph}
            \begin{description}
                \item[Input] A directed graph $G = (V, E)$ , $s\in V$ and $K \subseteq V \setminus \{s\}$.
                \item[Output] All strong $(s, K)$ cutsets in $G$.
                \item[Complexity] $O(|E|)$ time per cut.
                \item[Comment] An \textit{$(s, K)$-cut} is defined to be any cut $(X, Y)$ for which $s \in X$ and $K \cap Y \neq \emptyset$. A \textit{strong $(s, K)$ cutset} is minimal cuts of the form $(X, Y)$ where $s \in X$ and $K \subseteq Y$.
                \item[Reference] \cite{Provan1996}
            \end{description}
        \subsubsection{Enumeration of all quasi $(s, K)$ cuts in a graph}
            \begin{description}
                \item[Input] A graph $G = (V, E)$ , $s\in V$ and $K \subseteq V \setminus \{s\}$.
                \item[Output] All quasi $(s, K)$ cutsets in $G$.
                \item[Complexity] $O(|E|)$ time per cut.
                \item[Comment] An \textit{$(s, K)$-cut} is defined to be any cut $(X, Y)$ for which $s \in X$ and $K \cap Y \neq \emptyset$. A \textit{Quasi $(s, K)$ cut} is an edge set that is strong $(s, A)$-cutsets for some $A \subseteq K$ and $A \neq \emptyset$.
                \item[Reference] \cite{Provan1996}
            \end{description}
        \subsubsection{Enumeration of all quasi $(s, K)$ cuts in a directed graph}
            \begin{description}
                \item[Input] A directed graph $G = (V, E)$ , $s\in V$ and $K \subseteq V \setminus \{s\}$.
                \item[Output] All quasi $(s, K)$ cutsets in $G$.
                \item[Complexity] $O(|E|)$ time per cut.
                \item[Comment] An \textit{$(s, K)$-cut} is defined to be any cut $(X, Y)$ for which $s \in X$ and $K \cap Y \neq \emptyset$. A \textit{Quasi $(s, K)$ cut} is an edge set that is strong $(s, A)$-cutsets for some $A \subseteq K$ and $A \neq \emptyset$.
                \item[Reference] \cite{Provan1996}
            \end{description}
        \subsubsection{Enumeration of all $(s, K)$ cutsets in a graph}
            \begin{description}
                \item[Input] A graph $G = (V, E)$ , $s\in V$ and $K \subseteq V \setminus \{s\}$.
                \item[Output] All $(s, K)$ cutsets in $G$.
                \item[Complexity] $O(|E|)$ time per cut.
                \item[Comment] An \textit{$(s, K)$-cut} is defined to be any cut $(X, Y)$ for which $s \in X$ and $K \cap Y \neq \emptyset$. A \textit{$(s, K)$ cutset} is minimal $(s, K)$ cuts.
                \item[Reference] \cite{Provan1996}
            \end{description}
        \subsubsection{Enumeration of all minimal $a$-$b$ separators in a graph}
            \begin{description}
                \item[Input] An undirected connected simple graph $G = (V, E)$, non-adjacent vertices $a$ and $b$ in $G$.
                \item[Output] All minimal $a$-$b$ separator of $G$.
                \item[Complexity] $O(R_{ab}|V|^3)$ total time.
                \item[Comment] $R_{ab}$ is the number of minimal $a$-$b$ separators.
                \item[Reference] \cite{Shen1997}
            \end{description}
        \subsubsection{Enumeration of all minimal $a$-$b$ separators in a graph}
            \begin{description}
                \item[Input] An undirected connected simple graph $G = (V, E)$, non-adjacent vertices $a$ and $b$ in $G$.
                \item[Output] All minimal $a$-$b$ separator of $G$.
                \item[Complexity] $O(R_{ab}|V|/\log|V|)$ total time.
                \item[Comment] $R_{ab}$ is the number of minimal $a$-$b$ separators. This algorithm needs $O(|V|^3)$ processors on a CREW PRAM.
                \item[Reference] \cite{Shen1997}
            \end{description}
        \subsubsection{Enumeration of all minimal separators of a graph}
            \begin{description}
                \item[Input] A graph $G = (V, E)$.
                \item[Output] All minimal separators of $G$.
                \item[Complexity] $O(|V|^5R)$ total time.
                \item[Comment] $R$ is the number of solutions.
                \item[Reference] \cite{Kloks1998}
            \end{description}
        \subsubsection{Enumeration of all minimal separator of a graph}
            \begin{description}
                \item[Input] A graph $G = (V, E)$.
                \item[Output] All minimal separator of $G$.
                \item[Complexity] $O(|V|^3)$ time per solution.
                \item[Reference] \cite{Berry2000}
            \end{description}
        \subsubsection{Enumeration of all minimal separator of a chordal graph}
            \begin{description}
                \item[Input] A chordal graph $G = (V, E)$.
                \item[Output] All minimal separator of $G$.
                \item[Complexity] $O(|V|+|E|)$ total time.
                \item[Reference] \cite{Chandran2001}
            \end{description}
        \subsubsection{Enumeration of all cut conjunctions for a given set of vertex pairs in a graph}
            \begin{description}
                \item[Input] A graph $G = (V,E)$, and a collection $B = \{(s_1,t_1),...,(s_k,t_k)\}$ of $k$ vertex pairs $s_i,t_i \in V$.
                \item[Output] All cut conjunctions for $B$ in $G$.
                \item[Complexity] Incremental polynomial time.
                \item[Comment] A minimal edge sets $X \subseteq E$ is a \textit{cut conjunction} if, for all $i = 1, \dots, k$, vertices $s_i$ and $t_i$ is disconnected in $G' = (V, E\setminus X)$. A cut conjunction is a generalization of an $s-t$ cut.
                \item[Reference] \cite{Khachiyan2005}
            \end{description}
        \subsubsection{Enumeration of all minimal separators of a 3-connected planar graph}
            \begin{description}
                \item[Input] A 3-connected planar graph $G = (V, E)$.
                \item[Output] All minimal separators of $G$.
                \item[Complexity] $O(|V|)$ time per solution.
                \item[Reference] \cite{Mazoit2006}
            \end{description}
        \subsubsection{Enumeration of all cut conjunctions of a graph}
            \begin{description}
                \item[Input] A graph $G = (V, E)$ and a collection $B = \{(s_1, t_1), \dots, (s_k, t_k)\}$.
                \item[Output] All cut conjunctions of $G$.
                \item[Complexity] $O(K^2 \log(K) |V||E|^2 + K^2|B|(|V|+|E|)|E|^2)$ total time.
                \item[Comment] $K$ is the number of cut conjunctions.  $X$ is a cut conjunctions of $G$ if $X$ is a minimal edge set such that for all $i = 1, \dots, k$, a pair of vertices $s_i$ and $t_i$ is disconnected in $(V, E\setminus X)$.
                \item[Reference] \cite{Khachiyan2008a}
            \end{description}
        \subsubsection{Enumeration of all $(s, t)$-cuts in an weighted directed graph}
            \begin{description}
                \item[Input] An weighted directed graph $G = (V, E)$.
                \item[Output] All cuts in $G$ by non-decreasing weights ordering.
                \item[Complexity] $O(|V||E|\log(|V|^2/|E|))$ delay.
                \item[Reference] \cite{Yeh2010}
            \end{description}
        \subsubsection{Enumeration of all $(s, t)$-cuts in an weighted undirected graph}
            \begin{description}
                \item[Input] An weighted undirected graph $G = (V, E)$.
                \item[Output] All cuts in $G$ by non-decreasing weights ordering.
                \item[Complexity] $O(|V||E|\log(|V|^2/|E|))$ delay.
                \item[Reference] \cite{Yeh2010}
            \end{description}
     \subsection{Cycle}
        \subsubsection{Enumeration of all cycles in an $n$-cube, where $n \le 4$}
            \begin{description}
                \item[Input] An integer $n$. 
                \item[Output] All cycles (closed paths) in an $n$-cube. 
                \item[Reference] \cite{Gilbert1958}
            \end{description}
        \subsubsection{Enumeration of all cycles in a graph}
            \begin{description}
                \item[Input] A graph $G$. 
                \item[Output] All cycles in $G$. 
                \item[Reference] \cite{Welch1965}
            \end{description}
        \subsubsection{Enumeration of all simple cycles in a graph}
            \begin{description}
                \item[Input] A graph $G$.  
                \item[Output] All simple cycles in $G$. 
                \item[Complexity]  
                \item[Reference] \cite{Ponstein1966}
            \end{description}
        \subsubsection{Enumeration of all circuits in a graph}
            \begin{description}
                \item[Input] A graph $G$. 
                \item[Output] All circuits in $G$. 
                \item[Reference] \cite{Welch1966}
            \end{description}
        \subsubsection{Enumeration of all Hamiltonian circuits in a graph}
            \begin{description}
                \item[Input] A graph $G$. 
                \item[Output] All Hamiltonian circuits in $G$. 
                \item[Reference] \cite{Yau1967}
            \end{description}
        \subsubsection{Enumeration of all directed circuits in a directed graph}
            \begin{description}
                \item[Input] A directed graph $G$. 
                \item[Output] All directed circuits in $G$. 
                \item[Reference] \cite{Kamae1967}
            \end{description}
        \subsubsection{Enumeration of all cycles in a graph}
            \begin{description}
                \item[Input] A graph $G$. 
                \item[Output] All cycles in $G$. 
                \item[Reference] \cite{Danielson1968}
            \end{description}
        \subsubsection{Enumeration of all cycles in a graph}
            \begin{description}
                \item[Input] A graph $G$. 
                \item[Output] All cycles in $G$. 
                \item[Reference] \cite{Rao1969}
            \end{description}
        \subsubsection{Enumeration of all elementary circuit in a graph}
            \begin{description}
                \item[Input] A graph $G$.  
                \item[Output] All elementary circuit in $G$. 
                \item[Complexity]  
                \item[Reference] \cite{Tiernan1970}
            \end{description}
        \subsubsection{Enumeration of all cycles in a graph}
            \begin{description}
                \item[Input] A graph $G$. 
                \item[Output] All cycles in $G$. 
                \item[Reference] \cite{Char1970}
            \end{description}
        \subsubsection{Enumeration of all cycles in an undirected graph}
            \begin{description}
                \item[Input] An undirected graph $G$. 
                \item[Output] All cycles in $G$. 
                \item[Reference] \cite{Weinblatt1972}
            \end{description}
        \subsubsection{Enumeration of all cycles in a finite undirected graph}
            \begin{description}
                \item[Input] A finite undirected graph $G$. 
                \item[Output] All cycles in $G$. 
                \item[Comment] He claimed that J. T. Welch, Jr.'s algorithm is wrong. 
                \item[Reference] \cite{Weinblatt1972}
            \end{description}
        \subsubsection{Enumeration of all cycles in a directed graph}
            \begin{description}
                \item[Input] A directed graph $G=(V,E)$.
                \item[Output] All cycles in $G$.
                \item[Complexity] $O((|V|\cdot |E|)(C+1))$ total time and $O(|V| + |E|)$ space.
                \item[Comment] $C$ is the number of cycles included in $G$.
                \item[Reference] \cite{Tarjan1973}
            \end{description}
        \subsubsection{Enumeration of all cycles in a directed graph}
            \begin{description}
                \item[Input] A graph $G = (V, E)$. 
                \item[Output] All cycles in $G$. 
                \item[Complexity] $O(|E| + c(|V| \times |E|))$ total, where $c$ is the number of circuits in $G$. 
                \item[Reference] \cite{Ehrenfeucht1973}
            \end{description}
        \subsubsection{Enumeration of all cycles in a directed graph}
            \begin{description}
                \item[Input] A directed graph $G=(V, E)$.
                \item[Output] All cycles in $G$.
                \item[Complexity] $O((|V|+ |E|)(C+1))$ total time and $O(|V| + |E|)$ space.
                \item[Comment] $C$ is the number of cycles included in $G$.
                \item[Reference] \cite{Johnson1975}
            \end{description}
        \subsubsection{Enumeration of all cycles in a graph}
            \begin{description}
                \item[Input] A graph $G = (V, E)$.
                \item[Output] All cycles in $G$.
                \item[Complexity] $O(|E|)$ time per cycle with $O(|E|)$ space.
                \item[Reference] \cite{Read1975}
            \end{description}
        \subsubsection{Enumeration of all cycles in a directed graph}
            \begin{description}
                \item[Input] A directed graph $G = (V, E)$.
                \item[Output] All cycles in $G$.
                \item[Complexity] $O(|E|)$ time per cycle with $O(|E|)$ space.
                \item[Reference] \cite{Read1975}
            \end{description}
        \subsubsection{Enumeration of all cycles in a planar graph}
            \begin{description}
                \item[Input] A planar graph $G=(V,E)$.
                \item[Output] All cycles in $G$.
                \item[Complexity] $O(|V|(C+1))$ total time and $O(|V|)$ space.
                \item[Comment] $C$ is the number of cycles included in $G$.
                \item[Reference] \cite{Syso1981}
            \end{description}
        \subsubsection{Enumeration of all cycles of a given length in a graph}
            \begin{description}
                \item[Input] A graph $G$ and an integer $k$.
                \item[Output] All cycles of length $k$ in $G$.
                \item[Complexity] See paper.
                \item[Reference] \cite{Alon1997}
            \end{description}
        \subsubsection{Enumeration of all cycles in a graph}
            \begin{description}
                \item[Input] A graph $G$.
                \item[Output] All cycles in $G$.
                \item[Reference] \cite{Wild2008}
            \end{description}
        \subsubsection{Enumeration of all small cycles in a graph}
            \begin{description}
                \item[Input] A graph $G$.
                \item[Output] All cycles with length at most $5$ in $G$.
                \item[Reference] \cite{Wild2008}
            \end{description}
        \subsubsection{Enumeration of all chordless cycles in a graph}
            \begin{description}
                \item[Input] A graph $G$.
                \item[Output] All chordless cycles in $G$.
                \item[Reference] \cite{Wild2008}
            \end{description}
        \subsubsection{Enumeration of all Hamiltonian cycles in a graph}
            \begin{description}
                \item[Input] A graph $G$.
                \item[Output] All Hamiltonian cycles in $G$.
                \item[Reference] \cite{Wild2008}
            \end{description}
        \subsubsection{Enumeration of all cycles in a graph}
            \begin{description}
                \item[Input] A graph $G=(V,E)$.
                \item[Output] All cycles in $G$.
                \item[Complexity] $O(|E| + \sum_{c \in \mathcal{C}(G)}|c|)$ total time.
                \item[Comment] $\mathcal{C}(G)$ is the set of all cycles in $G$.
                \item[Reference] \cite{Ferreira2012}
            \end{description}
        \subsubsection{Enumeration of all chordless cycles in a graph}
            \begin{description}
                \item[Input] A graph $G=(V,E)$.
                \item[Output] All chordless cycles in $G$.
                \item[Complexity] $\tilde{O}(|E| +|V|\cdot C )$ total time.
                \item[Comment] $C$ is the number of chordless cycles in $G$.
                \item[Reference] \cite{Ferreira2014}
            \end{description}
        \subsubsection{Enumeration of all chordless cycles in a graph}
            \begin{description}
                \item[Input] A graph $G=(V,E)$.
                \item[Output] All chordless cycles in $G$.
                \item[Complexity] $O(|V|+|E|)$ time per chordless cycle.
                \item[Reference] \cite{Unoc}
            \end{description}
     \subsection{Dominating set}
        \subsubsection{Enumeration of all minimal dominating sets in a graph}
            \begin{description}
                \item[Input] A graph $G = (V, E)$.
                \item[Output] All minimal dominating sets in $G$.
                \item[Complexity] $O(1.7159^{|V|})$ total time.
                \item[Reference] \cite{Fomin2008}
            \end{description}
        \subsubsection{Enumeration of $z$ smallest weighted edge dominating sets in a graph}
            \begin{description}
                \item[Input] An weighted graph $G = (V, E)$, and positive integers $k$ and $z$. Each edge of $G$ has a positive weight.
                \item[Output] $z$ smallest weighted edge dominating sets in $G$.
                \item[Complexity] $O(5.6^{2k}k^4z^2 +4^{2k}k^3z|V|)$ total time.
                \item[Reference] \cite{Wang2009}
            \end{description}
        \subsubsection{Enumeration of all minimal dominating sets in a line graph}
            \begin{description}
                \item[Input] A line graph $G$.
                \item[Output] All minimal dominating sets in $G$.
                \item[Complexity] $O(||G||^5N^6)$ total time.
                \item[Comment] $||G||$ is the size of $G$ and $N$ is the number of minimal dominating sets in $G$.
                \item[Reference] \cite{Kante2012}
            \end{description}
        \subsubsection{Enumeration of all minimal dominating sets in a path graph or $(C_4, C_5, craw)$-free graph}
            \begin{description}
                \item[Input] A line graph or $(C_4, C_5, craw)$-free graph $G$.
                \item[Output] All minimal dominating sets in $G$.
                \item[Complexity] $O(||G||^2N^3)$ total time.
                \item[Comment] $||G||$ is the size of $G$ and $N$ is the number of minimal dominating sets in $G$.
                \item[Reference] \cite{Kante2012}
            \end{description}
        \subsubsection{Enumeration of all minimal edge-dominating sets in a graph}
            \begin{description}
                \item[Input] A graph $G$.
                \item[Output] All minimal edge-dominating sets in $G$.
                \item[Complexity] $O(||L(G)||^5N^6)$ total time.
                \item[Comment] $L(G)$ is the line graph of $G$, $||L(G)||$ is the size of $L(G)$, and $N$ is the number of minimal edge-dominating sets in $G$.
                \item[Reference] \cite{Kante2012}
            \end{description}
        \subsubsection{Enumeration of all minimal dominating sets in an undirected permutation graph}
            \begin{description}
                \item[Input] An undirected permutation graph $G=(V, E)$.
                \item[Output] All minimal dominating sets.
                \item[Complexity] $O(|V|)$ delay with $O(|V|^8)$ pre-processing.
                \item[Reference] \cite{Kante2013}
            \end{description}
        \subsubsection{Enumeration of all minimal dominating sets in an undirected interval graph}
            \begin{description}
                \item[Input] An undirected interval graph $G=(V, E)$.
                \item[Output] All minimal dominating sets.
                \item[Complexity] $O(|V|)$ delay with $O(|V|^3)$ pre-processing.
                \item[Reference] \cite{Kante2013}
            \end{description}
        \subsubsection{Enumeration of all minimal edge dominating sets in a graph}
            \begin{description}
                \item[Input] A graph $G = (V, E)$.
                \item[Output] All minimal edge dominating sets in $G$.
                \item[Complexity] $O(|V|^6|\mathcal{L}|)$ delay.
                \item[Comment] $\mathcal{L}$ is the set of already generated solutions.
                \item[Reference] \cite{Golovach2013a}
            \end{description}
        \subsubsection{Enumeration of all minimal dominating sets in a line graph}
            \begin{description}
                \item[Input] A line graph $G = (V, E)$.
                \item[Output] All minimal dominating sets in $G$.
                \item[Complexity] $O(|V|^2|E|^2|\mathcal{L}|)$ delay.
                \item[Comment] $\mathcal{L}$ is the set of already generated solutions.
                \item[Reference] \cite{Golovach2013a}
            \end{description}
        \subsubsection{Enumeration of all minimal edge dominating sets in a bipartite graph}
            \begin{description}
                \item[Input] A bipartite graph $G = (V, E)$.
                \item[Output] All minimal edge dominating sets in $G$.
                \item[Complexity] $O(|V|^4|\mathcal{L}|)$ delay.
                \item[Comment] $\mathcal{L}$ is the set of already generated solutions.
                \item[Reference] \cite{Golovach2013a}
            \end{description}
        \subsubsection{Enumeration of all minimal dominating sets in the line graph of a bipartite graph}
            \begin{description}
                \item[Input] A graph $G = (V, E)$.
                \item[Output] All minimal dominating sets in $G$.
                \item[Complexity] $O(|V|^2|E||\mathcal{L}|)$ delay.
                \item[Comment] $\mathcal{L}$ is the set of already generated solutions.
                \item[Reference] \cite{Golovach2013a}
            \end{description}
        \subsubsection{Enumeration of all minimal dominating sets in a graph of girth at least 7}
            \begin{description}
                \item[Input] A graph $G = (V, E)$ of girth at least 7.
                \item[Output] All minimal dominating sets in $G$.
                \item[Complexity] $O(|V|^2|E||\mathcal{L}|^2)$ delay.
                \item[Comment] $\mathcal{L}$ is the set of already generated solutions.
                \item[Reference] \cite{Golovach2013a}
            \end{description}
        \subsubsection{Enumeration of all 2-dominating sets in a tree}
            \begin{description}
                \item[Input] A tree $T = (V, E)$.
                \item[Output] All 2-dominating sets of $T$.
                \item[Complexity] $O(1.3248^n)$ total time.
                \item[Comment] If a subset $U\subseteq V$ is a 2-dominating set if every vertex $v \in V \setminus U$ has at least two neighbors in $U$.
                \item[Reference] \cite{Krzywkowski2013}
            \end{description}
        \subsubsection{Enumeration of all minimal dominating sets in a $P_6$-free chordal graph}
            \begin{description}
                \item[Input] A $P_6$-free chordal graph $G = (V, E)$.
                \item[Output] All minimal dominating sets in $G$.
                \item[Complexity] Linear delay with $O(|V|^2)$ space.
                \item[Reference] \cite{Kante2014}
            \end{description}
        \subsubsection{Enumeration of all minimal dominating sets in a chordal bipartite graph}
            \begin{description}
                \item[Input] A chordal bipartite graph $G$. 
                \item[Output] All minimal dominating sets in $G$. 
                \item[Complexity] $O(n^3m|\mathcal{L}|^2)$ delay and the total running time is $O(n^3m|\mathcal{L}^*|^2)$. 
                \item[Comment] $n$ is the number vertices in $G$, 
            $m$ is the number of edges in $G$, 
            $\mathcal{L}$ is the family of already generated minimal dominating sets, and 
            $\mathcal{L}^*$ is the family of all minimal dominating sets. 
        
                \item[Reference] \cite{Golovach2015}
            \end{description}
     \subsection{Drawing}
        \subsubsection{Enumeration of all rectangle drawings with $n$ faces}
            \begin{description}
                \item[Input] An integer $n$.
                \item[Output] All rectangle drawings with $n$ faces.
                \item[Complexity] $O(n)$ time per drawing and space.
                \item[Reference] \cite{Takagi2004}
            \end{description}
     \subsection{Feedback arc set}
        \subsubsection{Enumeration of all minimal feedback arc sets in a graph}
            \begin{description}
                \item[Input] A graph $G$. 
                \item[Output] All minimal feedback arc sets in $G$. 
                \item[Reference] \cite{Yau1967}
            \end{description}
        \subsubsection{Enumeration of all minimal feedback arc sets in a directed graph}
            \begin{description}
                \item[Input] A directed graph $G=(V,E)$.
                \item[Output] All minimal feedback arc sets in $G$.
                \item[Complexity] $O(|V||E|(|V|+|E|))$ time delay.
                \item[Reference] \cite{Schwikowski2002}
            \end{description}
     \subsection{Feedback vertex set}
        \subsubsection{Enumeration of all minimal feedback vertex sets in a graph}
            \begin{description}
                \item[Input] A graph $G$. 
                \item[Output] All minimal feedback vertex sets in $G$. 
                \item[Reference] \cite{Yau1967}
            \end{description}
        \subsubsection{Enumeration of all feedback vertex sets in a strongly connected directed graph}
            \begin{description}
                \item[Input] A strongly connected directed graph $G = (V, E)$ and an integer $k$.
                \item[Output] All feedback vertex sets of size $k$ in $G$.
                \item[Complexity] $O(|V|^{k-1}|E|)$ total time and $O(|E|)$ space.
                \item[Reference] \cite{Garey1978}
            \end{description}
        \subsubsection{Enumeration of all minimal feedback vertex sets in a graph}
            \begin{description}
                \item[Input] A graph $G=(V,E)$.
                \item[Output] All minimal feedback vertex sets in $G$.
                \item[Complexity] $O(|V||E|(|V|+|E|))$ time delay.
                \item[Reference] \cite{Schwikowski2002}
            \end{description}
        \subsubsection{Enumeration of all minimal feedback vertex sets in a directed graph}
            \begin{description}
                \item[Input] A directed graph $G=(V,E)$.
                \item[Output] All minimal feedback vertex sets in $G$.
                \item[Complexity] $O(|V|^2(|V|+|E|))$ time delay.
                \item[Reference] \cite{Schwikowski2002}
            \end{description}
     \subsection{General}
        \subsubsection{Enumeration of all graphs in an almost sure first order fimiles}
            \begin{description}
                \item[Input] A first order language $\theta$.
                \item[Output] All graphs in $\mathcal{G}_\theta$.
                \item[Complexity] Polynomial space and delay.
                \item[Reference] \cite{Goldberg1993b}
            \end{description}
     \subsection{Independent set}
        \subsubsection{Enumeration of all maximal independent sets in a chordal graph}
            \begin{description}
                \item[Input] A chordal graph $G=(V,E)$.
                \item[Output] All maximal independent sets in $G$.
                \item[Complexity] $O(|V||E|\mu)$ total time.
                \item[Comment] $\mu$ is the number of maximal independent sets of $G$. This algorithm can also enumerate all the maximal cliques.
                \item[Reference] \cite{Tsukiyama1977a}
            \end{description}
        \subsubsection{Enumeration of all maximal independent sets in a claw-free graph}
            \begin{description}
                \item[Input] A claw-free graph $G$.
                \item[Output] All maximal independent sets in $G$.
                \item[Reference] \cite{Minty1980}
            \end{description}
        \subsubsection{Enumeration of all maximal independent sets in an undirected graph}
            \begin{description}
                \item[Input] A graph $G$.
                \item[Output] All maximal independent sets in $G$ in lexicographically.
                \item[Reference] \cite{Loukakis1981}
            \end{description}
        \subsubsection{Enumeration of all maximal independent sets in an interval graph}
            \begin{description}
                \item[Input] An interval graph $G = (V, E)$.
                \item[Output] All maximal independent sets in $G$.
                \item[Complexity] $O(|V|^2 + \beta)$ total time.
                \item[Comment] $\beta$ is the sum of the vertices in all maximal independent sets of $G$.
                \item[Reference] \cite{Leung1984}
            \end{description}
        \subsubsection{Enumeration of all maximal independent sets in a circular-arc graph}
            \begin{description}
                \item[Input] A circular-arc graph $G = (V, E)$.
                \item[Output] All maximal independent sets in $G$.
                \item[Complexity] $O(|V|^2 + \beta)$ total time.
                \item[Comment] $\beta$ is the sum of the vertices in all maximal independent sets of $G$.
                \item[Reference] \cite{Leung1984}
            \end{description}
        \subsubsection{Enumeration of all maximal independent sets in a chordal graph}
            \begin{description}
                \item[Input] A chordal graph $G = (V, E)$.
                \item[Output] All maximal independent sets in $G$.
                \item[Complexity] $O((|V| + |E|)N)$ total time.
                \item[Comment] $N$ is the number of solutions.
                \item[Reference] \cite{Leung1984}
            \end{description}
        \subsubsection{Enumeration of all maximal independent sets in a graph}
            \begin{description}
                \item[Input] A graph $G=(V,E)$.
                \item[Output] All maximal independent sets included in $G$.
                \item[Complexity] $O(n(m+n\log C))=O(n^3)$ delay and exponential space.
                \item[Comment] $C$: total number of maximal independent sets, $n$: total number of vertices, and $m$: total number of edges. Is there no polynomial space and delay algorithm?
                \item[Reference] \cite{Johnson1988}
            \end{description}
        \subsubsection{Enumeration of all maximum independent sets of a bipartite graph}
            \begin{description}
                \item[Input] A bipartite graph $B =(V, E)$.
                \item[Output] All maximum independent sets of $B$.
                \item[Complexity] $O(|V|^{2.5} + N)$.
                \item[Comment] $N$ is the number of maximum independent sets of $B$.
                \item[Reference] \cite{Kashiwabara1992}
            \end{description}
        \subsubsection{Enumeration of all maximal independent sets on a tree in lexicographic order}
            \begin{description}
                \item[Input] Tree $T = (V, E)$.
                \item[Output] All maximal independent sets on $T$ in lexicographic order.
                \item[Complexity] $O(|V|^2)$ delay with $O(|V|)$ space.
                \item[Reference] \cite{Chang1994a}
            \end{description}
        \subsubsection{Enumeration of all maximum independent set of a graph}
            \begin{description}
                \item[Input] A graph $G = (V, E)$.
                \item[Output] All maximum independent set of $G$.
                \item[Complexity] $O(2^{0.114|E|})$ total time and polynomial space.
                \item[Reference] \cite{Beigel1999}
            \end{description}
        \subsubsection{Enumeration of all maximum independent set of a graph}
            \begin{description}
                \item[Input] A graph $G = (V, E)$ with maximum degree 3.
                \item[Output] All maximum independent set of $G$.
                \item[Complexity] $O(2^{0.171|V|})$ total time and polynomial space.
                \item[Reference] \cite{Beigel1999}
            \end{description}
        \subsubsection{Enumeration of all maximum independent set of a graph}
            \begin{description}
                \item[Input] A graph $G = (V, E)$ with maximum degree 4.
                \item[Output] All maximum independent set of $G$.
                \item[Complexity] $O(2^{0.228|V|})$ total time and polynomial space.
                \item[Reference] \cite{Beigel1999}
            \end{description}
        \subsubsection{Enumeration of all maximum independent set of a graph}
            \begin{description}
                \item[Input] A graph $G = (V, E)$.
                \item[Output] All maximum independent set of $G$.
                \item[Complexity] $O(2^{0.290|V|})$ total time and polynomial space.
                \item[Reference] \cite{Beigel1999}
            \end{description}
        \subsubsection{Enumeration of all maximal independent sets of a graph}
            \begin{description}
                \item[Input] A graph $G = (V, E)$ and a position integer $k$.
                \item[Output] Enumeration of all maximal independent sets with at most size $k$ of $G$.
                \item[Complexity] $O(3^{4k-|V|} 4^{|V|-3k})$ total time.
                \item[Reference] \cite{Eppstein2003a}
            \end{description}
        \subsubsection{Enumeration of all independent sets in a chordal graph}
            \begin{description}
                \item[Input] A chordal graph $G = (V, E)$.
                \item[Output] All independent sets in $G$.
                \item[Complexity] Constant time per solution on average after $O(|V| + |E|)$ time for preprocessing.
                \item[Reference] \cite{Okamoto2005a}
            \end{description}
        \subsubsection{Enumeration of all independent sets of size $k$ in a chordal graph}
            \begin{description}
                \item[Input] A chordal graph $G = (V, E)$ and a positive integer $k$.
                \item[Output] All independent sets of size $k$ in $G$.
                \item[Complexity] Constant time per solution on average after $O((|V| + |E|)|V|^2)$ time for preprocessing.
                \item[Reference] \cite{Okamoto2005a}
            \end{description}
        \subsubsection{Enumeration of all maximum independent sets in a chordal graph}
            \begin{description}
                \item[Input] A chordal graph $G = (V, E)$.
                \item[Output] All maximum independent sets in $G$.
                \item[Complexity] Constant time per solution on average after $O((|V| + |E|)|V|^2)$ time for preprocessing.
                \item[Reference] \cite{Okamoto2005a}
            \end{description}
        \subsubsection{Enumeration of all independent sets in an input chordal graph}
            \begin{description}
                \item[Input] A chordal graph $G=(V,E)$.
                \item[Output] All independent sets in $G$.
                \item[Complexity] $O(1)$ delay and $O(|V|(|V|+|E|))$ time and space for preprocessing.
                \item[Comment] Counting all independent sets in an input chordal graph needs $O(n+m)$ time.
                \item[Reference] \cite{Okamoto2008}
            \end{description}
        \subsubsection{Enumeration of all maximum independent sets in an input chordal graph}
            \begin{description}
                \item[Input] A chordal graph $G=(V,E)$.
                \item[Output] All maximum independent sets in $G$.
                \item[Complexity] $O(1)$ delay and $O(|V|(|V|+|E|))$ time and space for preprocessing.
                \item[Comment] Counting all maximum independent sets in an input chordal graph needs $O(n+m)$ time.
                \item[Reference] \cite{Okamoto2008}
            \end{description}
        \subsubsection{Enumeration of all independent sets with $k$ vertices in an input chordal graph}
            \begin{description}
                \item[Input] A chordal graph $G=(V,E)$ and an integer $k$.
                \item[Output] All maximum independent sets with $k$ vertices in $G$.
                \item[Complexity] $O(1)$ delay and $O(k|V|(|V|+|E|))$ time and space for preprocessing.
                \item[Comment] The number of independent sets with $k$ vertices in an input chordal graph needs $O(k^2(|V|+|E|))$ time.
                \item[Reference] \cite{Okamoto2008}
            \end{description}
     \subsection{Interval graph}
        \subsubsection{Enumeration of all interval supergraph that contains a given graph}
            \begin{description}
                \item[Input] A graph $G = (V, E)$.
                \item[Output] All interval supergraphs each of which contain $G$.
                \item[Complexity] $O(|V|^3)$ time for each and $O(|V|^2)$ space.
                \item[Reference] \cite{Kiyomi2006a}
            \end{description}
        \subsubsection{Enumeration of all interval graph of a given graph}
            \begin{description}
                \item[Input] A graph $G = (V, E)$.
                \item[Output] All interval graphs of $G$.
                \item[Complexity] $O((|V|+|E|)^2)$ time for each.
                \item[Reference] \cite{Kiyomi2006a}
            \end{description}
        \subsubsection{Enumeration of Proper Interval Graphs}
            \begin{description}
                \item[Input] An integer $n$.
                \item[Output] All proper interval graphs with $n$ vertices.
                \item[Complexity] $O(1)$ time per proper interval graph and $O(n)$ space, after $O(n^2)$ preprocessing time.
                \item[Comment] Preprocessing: generating the complete graph with $n$ vertices.
                \item[Reference] \cite{Saitoh2010a}
            \end{description}
     \subsection{Matching}
        \subsubsection{Enumeration of the $k$ best perfect matchings of a graph}
            \begin{description}
                \item[Input] A graph $G = (V, E)$ and an integer $k$.
                \item[Output] The $k$ best perfect matchings of $G$ in order.
                \item[Complexity] $O(k|V|^3)$ total time.
                \item[Reference] \cite{Chegireddy1987}
            \end{description}
        \subsubsection{Enumeration of all stable marriage}
            \begin{description}
                \item[Input] A graph $G = (V, E)$.
                \item[Output] All stable marriage of $G$.
                \item[Complexity] $O(|V|)$ time per solution and $O(|V|^2)$ space.
                \item[Reference] \cite{Gusfield1989}
            \end{description}
        \subsubsection{Enumeration of all minimum cost perfect matchings in an weighted bipartite graph}
            \begin{description}
                \item[Input] An weighted bipartite graph $B = (V, E)$.
                \item[Output] All minimum cost perfect matchings in $B$.
                \item[Complexity] $O(|V|(|V| + |E|))$ time per solution and $O(|V| + |E|)$ space
                \item[Reference] \cite{Fukuda1992}
            \end{description}
        \subsubsection{Enumeration of all perfect matchings in a bipartite graph}
            \begin{description}
                \item[Input] A bipartite graph $B = (U, V, E)$, where $|U| = |V|$.
                \item[Output] All perfect matchings in $B$.
                \item[Complexity] $O(c(|V| + |E|))$ total time and $O(|V| + |E|)$ space, after $O(n^{2.5})$ preprocessing time.
                \item[Comment] $c$ is the number of solutions.
                \item[Reference] \cite{Fukuda1994}
            \end{description}
        \subsubsection{Enumeration of the $k$ best perfect matchings of a graph}
            \begin{description}
                \item[Input] A graph $G = (V, E)$ and an integer $k$.
                \item[Output] The $k$ best perfect matchings of $G$ in decreasing order.
                \item[Complexity] $O(k|V|^3)$ total time with $O(k|V|^2)$ space.
                \item[Reference] \cite{Matsui1994}
            \end{description}
        \subsubsection{Enumeration of all perfect, maximum, and maximal matchings in bipartite graphs}
            \begin{description}
                \item[Input] A bipartite graph $B=(V, E)$.
                \item[Output] All perfect, maximum, and maximal matching in $B$.
                \item[Complexity] $O(|V|)$ time per matching.
                \item[Reference] \cite{Uno1997}
            \end{description}
        \subsubsection{Enumeration of all maximal matchings in a graph}
            \begin{description}
                \item[Input] A graph $G = (V, E)$.
                \item[Output] All maximal matchings in $G$.
                \item[Complexity] $O(|V| + |E| + \Delta N)$ total time and $O(|V| + |E|)$ space.
                \item[Comment] $N$ is the number of solutions and $\Delta$ is the maximum degree in $G$.
                \item[Reference] \cite{Uno2001}
            \end{description}
        \subsubsection{Enumeration of all minimal blocker in a bipartite graph}
            \begin{description}
                \item[Input] A bipartite graph $G = (U, V, E)$.
                \item[Output] All minimal blocker in $G$.
                \item[Complexity] Polynomial delay and space.
                \item[Comment] A \textit{blocker} of $G$ is an edge subset $X$ of $E$ such that $G' = (U, V, E \setminus X)$ has no perfect matching.
                \item[Reference] \cite{Boros2006}
            \end{description}
        \subsubsection{Enumeration of all basic perfect 2-matchings in a graph}
            \begin{description}
                \item[Input] A graph $G = (V, E)$.
                \item[Output] All basic perfect 2-matchings in $G$.
                \item[Complexity] Incremental polynomial delay.
                \item[Comment] A \textit{basic 2-matching} of $G$ is a subset of edges that cover the vertices with vertex-disjoint edges and vertex-disjoint odd cycles.
                \item[Reference] \cite{Boros2006}
            \end{description}
        \subsubsection{Enumeration of all $d$-factor in a bipartite graph}
            \begin{description}
                \item[Input] A bipartite graph $G = (U, V, E)$ and any non negative function $d: A \cup B \to \{0, 1, \cdots, |U| + |V|\}$.
                \item[Output] All $d$-factor in $G$.
                \item[Complexity] $O(|E|)$ delay.
                \item[Comment] A $d$-factor in $G$ is a subgraph $G' = (U, V, X)$ covering all vertices of $G$, whose each vertex $v$ has degree $d(v)$. If for any $v \in U \cup V$, $d(v) = 1$, $G'$ is a perfect matching.
                \item[Reference] \cite{Boros2006}
            \end{description}
        \subsubsection{Enumeration of all maximal induced matchings in a triangle-free graph}
            \begin{description}
                \item[Input] A triangle-free graph $G$. 
                \item[Output] All maximal induced matchings in $G$. 
                \item[Complexity] $O(1.4423^n)$ total time with polynomial delay. 
                \item[Comment] $n$ is the number of vertices in $G$. 
                \item[Reference] \cite{Basavaraju2014}
            \end{description}
     \subsection{Matroid}
        \subsubsection{Enumeration of all bases of a graphic matroid in a graph}
            \begin{description}
                \item[Input] A graph $G = (V, E)$.
                \item[Output] All bases of a graphic matroid in $G$.
                \item[Complexity] $O(|V| + |E| + N)$ total time and $O(|V| + |E|)$ space.
                \item[Comment] $N$ is the number of solutions.  If $G$ is connected, any base is a spanning tree.
                \item[Reference] \cite{Uno1998}
            \end{description}
        \subsubsection{Enumeration of all bases of a linear matroid in a graph}
            \begin{description}
                \item[Input] A graph $G = (V, E)$.
                \item[Output] All bases of a linear matroid in $G$.
                \item[Complexity] $O(|V|)$ time per solution and $O(|V|^2|E|)$ preprocessing after time.
                \item[Reference] \cite{Uno1998}
            \end{description}
        \subsubsection{Enumeration of all bases of a matching matroid in a graph}
            \begin{description}
                \item[Input] A graph $G = (V, E)$.
                \item[Output] All bases of a matching matroid in $G$.
                \item[Complexity] $O(|V| + |E|)$ time per solution.
                \item[Reference] \cite{Uno1998}
            \end{description}
     \subsection{Ordering}
        \subsubsection{Enumeration of all topological sortings of a directed graph}
            \begin{description}
                \item[Input] A directed graph $G =(V, E)$.
                \item[Output] All topological sortings of $G$.
                \item[Complexity] $O(|V| + |E|)$ time per sorting and $O(|V| + |E|)$ space.
                \item[Reference] \cite{Knuth1974}
            \end{description}
        \subsubsection{Enumeration of all topological sortings of a given set in lexicographically}
            \begin{description}
                \item[Input] An $n$-element set $S$. 
                \item[Output] All topological sortings of $S$ in lexicographically.
                \item[Complexity] $O(m)$ time per solution(?). 
                \item[Reference] \cite{Knuth1979}
            \end{description}
        \subsubsection{Enumeration of all topological sortings of a po set}
            \begin{description}
                \item[Input] A partial order set $P$.
                \item[Output] All topological sortings of $P$.
                \item[Complexity] $O(|P|)$ time per solution.
                \item[Comment] $|P|$ is the number of objects in $P$.
                \item[Reference] \cite{Varol1981}
            \end{description}
        \subsubsection{Enumeration of all topological sortings in a directed acyclic graph}
            \begin{description}
                \item[Input] A directed graph $G$.
                \item[Output] All topological sortings in $G$.
                \item[Complexity] $O(1)$ amortized time per solution with $O(|G|)$.
                \item[Comment] Linear extensions correspond to topological sortings.
                \item[Reference] \cite{Pruesse1991a}
            \end{description}
        \subsubsection{Enumeration of all topological sortings}
            \begin{description}
                \item[Input] A graph $G$.
                \item[Output] All topological sortings in $G$.
                \item[Complexity] $O(1)$ amortized time per solution.
                \item[Comment] A topological sorting is also known as a linear extension.
                \item[Reference] \cite{Ruskey1992a}
            \end{description}
        \subsubsection{Enumeration of all topological sortings}
            \begin{description}
                \item[Input] A directed acyclic graph $D$.
                \item[Output] All topological sortings $D$.
                \item[Complexity] $O(1)$ amortized time per topological sorting and $O(|V|)$ space in addition to the space used for $D$.
                \item[Comment] Linear sortings correspond to topological sortings.
                \item[Reference] \cite{Pruesse1994}
            \end{description}
        \subsubsection{Enumeration of all linear extensions of a given poset}
            \begin{description}
                \item[Input] A poset $P$.
                \item[Output] All linear extensions of $P$.
                \item[Complexity] $O(1)$ time per solution.
                \item[Comment] Their algorithm is a loop-free algorithm.
                \item[Reference] \cite{Canfield1995}
            \end{description}
        \subsubsection{Enumeration of all topological sortings of an acyclic directed graph}
            \begin{description}
                \item[Input] An acyclic directed graph $G = (V, E)$.
                \item[Output] All topological sortings of $G$.
                \item[Complexity] $O(|V|N)$ total time and $O(|V||E|)$ space.
                \item[Comment] $N$ is the number of solutions.
                \item[Reference] \cite{Avis1996}
            \end{description}
        \subsubsection{Enumeration of all perfect elimination orderings}
            \begin{description}
                \item[Input] A chordal graph $G$.
                \item[Output] All perfect elimination orderings of $G$.
                \item[Complexity] Constant amortized time per solution.
                \item[Reference] \cite{Chandran2003}
            \end{description}
        \subsubsection{Enumeration of all forest extensions of a partially ordered set}
            \begin{description}
                \item[Input] A partially ordered set $P$.
                \item[Output] All forest extensions of $P$.
                \item[Complexity] $O(|E|^2)$ delay and $O(|E||R|)$ space.
                \item[Comment] $E$ is the set of elements. $R$ is the binary relation on $E$.
                \item[Reference] \cite{Szwarcfiter2003b}
            \end{description}
        \subsubsection{Enumeration of all topological sortings of a directed acyclic graph}
            \begin{description}
                \item[Input] A directed acyclic graph $D$.
                \item[Output] All topological sortings $D$.
                \item[Complexity] $O(1)$ delay per topological sorting.
                \item[Comment] Linear extensions correspond to topological sortings.
                \item[Reference] \cite{Ono2005}
            \end{description}
        \subsubsection{Enumeration of all realizer of a triangulated planar graph}
            \begin{description}
                \item[Input] A triangulated planar graph $G = (V, E)$.
                \item[Output] All realizer of $G$.
                \item[Complexity] $O(|V|)$ time per realizer.
                \item[Reference] \cite{Yamanaka2006}
            \end{description}
        \subsubsection{Enumeration of all perfect elimination orderings of a chordal graph}
            \begin{description}
                \item[Input] A chordal graph $G = (V, E)$.
                \item[Output] All perfect elimination orderings of $G$.
                \item[Complexity] $O(1)$ time per solution on average with $O(|V|^2)$ space and $O(|V|^3)$ with $O(|V|^2)$ space pre-computation.
                \item[Reference] \cite{Matsui2008}
            \end{description}
        \subsubsection{Enumeration of all perfect sequences in a chordal graph}
            \begin{description}
                \item[Input] A chordal graph $G = (V, E)$.
                \item[Output] All perfect sequences of $G$.
                \item[Complexity] $O(1)$ time per graph with $O(|V|^2)$ space with $O(|V|^3)$ time and $O(|V|^2)$ space pre-computation.
                \item[Reference] \cite{Matsui2010}
            \end{description}
     \subsection{Orientation}
        \subsubsection{Enumeration of all acyclic orientation of a graph}
            \begin{description}
                \item[Input] A graph $G = (V, E)$.
                \item[Output] All acyclic orientation of $G$.
                \item[Complexity] $O(N(|V|+|E|))$ total time ($O(|V|(|V|+E|))$ delay) and $O(|V| + |E|)$ space.
                \item[Comment] A acyclic orientation of $G$ is an assignment of directions of each edge such that $G$ is acyclic.
                \item[Reference] \cite{Barbosa1999}
            \end{description}
        \subsubsection{Enumeration of all $(s, t)$-orientations of a biconnected planar graph}
            \begin{description}
                \item[Input] A biconnected planar graph $G = (V, E)$ and an edge $(s, t)$ in $G$.
                \item[Output] All $(s, t)$-orientations of $G$.
                \item[Complexity] $O(|V|)$ time per solution.
                \item[Reference] \cite{Setiawan2011}
            \end{description}
     \subsection{Other}
        \subsubsection{Enumeration of all Hamiltonian centers in a graph}
            \begin{description}
                \item[Input] A graph $G$. 
                \item[Output] All Hamiltonian centers in $G$. 
                \item[Reference] \cite{Yau1967}
            \end{description}
        \subsubsection{Enumeration of all CA-sets of a directed graph}
            \begin{description}
                \item[Input] Directed graph $G = (V, E)$.
                \item[Output] All CA-sets of $G$.
                \item[Complexity] $O(|V|^{2.49+} + \gamma)$.
                \item[Comment] $\gamma$ is the output size. $S \subset V$ is a \textit{CA-set} if, for each $v \in S$, all ancestor of $v$ belongs to $S$.
                \item[Reference] \cite{Kashiwabara1992}
            \end{description}
        \subsubsection{Enumeration of all maximal induced subgraphs for (connected) hereditary graph properties}
            \begin{description}
                \item[Input] A graph $G$.
                \item[Output] All maximal induced subgraphs in $\mathcal{P}(G)$.
                \item[Complexity] See the paper.
                \item[Comment] $\mathcal{P}$ is a set of subgraphs of $G$ with (connected) hereditary graph properties.
                \item[Reference] \cite{Cohen2008}
            \end{description}
     \subsection{Path}
        \subsubsection{Enumeration of all simple paths in a graph}
            \begin{description}
                \item[Input] A graph $G$.  
                \item[Output] All simple paths in $G$. 
                \item[Complexity]  
                \item[Reference] \cite{Ponstein1966}
            \end{description}
        \subsubsection{Enumeration of all Hamiltonian paths in a graph}
            \begin{description}
                \item[Input] A graph $G$. 
                \item[Output] All Hamiltonian paths in $G$. 
                \item[Reference] \cite{Yau1967}
            \end{description}
        \subsubsection{Enumeration of all directed paths in a directed graph}
            \begin{description}
                \item[Input] A directed graph $G$. 
                \item[Output] All directed paths in $G$. 
                \item[Reference] \cite{Kamae1967}
            \end{description}
        \subsubsection{Enumeration of all paths in a graph}
            \begin{description}
                \item[Input] A graph $G$. 
                \item[Output] All paths in $G$. 
                \item[Reference] \cite{Kroft1967b}
            \end{description}
        \subsubsection{Enumeration of all paths in a graph}
            \begin{description}
                \item[Input] A graph $G$. 
                \item[Output] All paths in $G$. 
                \item[Reference] \cite{Danielson1968}
            \end{description}
        \subsubsection{Enumeration of $k$ shortest paths in a graph}
            \begin{description}
                \item[Input] A graph $G = (V, E)$.
                \item[Output] $K$ shortest paths in $G$.
                \item[Complexity] $O(|V|^3)$ total time.
                \item[Reference] \cite{Yen1971}
            \end{description}
        \subsubsection{Enumeration of $k$ shortest paths in a graph}
            \begin{description}
                \item[Input] A graph $G = (V, E)$.
                \item[Output] $k$ shortest paths in $G$.
                \item[Complexity] $O(k|V|c(|V|))$ total time.
                \item[Comment] (?) $c(n)$ is the time complexity to find an optimal solution to a problem with $n$ (0, 1) variables.
                \item[Reference] \cite{Lawler1972}
            \end{description}
        \subsubsection{Enumeration of all paths in a graph}
            \begin{description}
                \item[Input] A graph $G = (V, E)$.
                \item[Output] All paths in $G$.
                \item[Complexity] $O(|E|)$ time per path with $O(|E|)$ space.
                \item[Reference] \cite{Read1975}
            \end{description}
        \subsubsection{Enumeration of all paths in a directed graph}
            \begin{description}
                \item[Input] A directed graph $G = (V, E)$.
                \item[Output] All paths in $G$.
                \item[Complexity] $O(|E|)$ time per path with $O(|E|)$ space.
                \item[Reference] \cite{Read1975}
            \end{description}
        \subsubsection{Enumeration of $k$ shortest paths in a graph}
            \begin{description}
                \item[Input] A graph $G = (V, E). $
                \item[Output] $K$ shortest paths in $G$.
                \item[Complexity] $O(k|V|^3)$ total time.
                \item[Reference] \cite{Shier1979}
            \end{description}
        \subsubsection{Generation of the $k$-th longest path in a tree}
            \begin{description}
                \item[Input] A tree $T = (V, E)$ and an integer $k$.
                \item[Output] The $k$-th longest path in $T$.
                \item[Complexity] $O(n\log^2 n)$ time.
                \item[Reference] \cite{Megiddo1981}
            \end{description}
        \subsubsection{Enumeration of all shortest paths in a graph}
            \begin{description}
                \item[Input] A graph $G$.
                \item[Output] All shortest paths in $G$.
                \item[Reference] \cite{Florian1981}
            \end{description}
        \subsubsection{Enumeration of $k$ shortest paths of a directed graph}
            \begin{description}
                \item[Input] A graph $G = (V, E)$.
                \item[Output] $k$ shortest paths that may contains cycles in $G$.
                \item[Reference] \cite{Martins1984}
            \end{description}
        \subsubsection{Enumeration of all quickest paths in a network}
            \begin{description}
                \item[Input] A network $N = (V, E, c, \ell)$.
                \item[Output] All quickest paths in $N$.
                \item[Complexity] $O(rS|V||E| + rS|V|^2\log|V|)$ total time.
                \item[Comment] $c$ is a positive edge weight function and $\ell$ is a nonnegative edge weight function. $r$ is the number of distinct capacity value of $N$. $S$ is the number of solutions.
                \item[Reference] \cite{Rosen1991}
            \end{description}
        \subsubsection{Counting all acyclic walks in a graph}
            \begin{description}
                \item[Input] A graph $G=(V,E)$.
                \item[Output] The number of acyclic walks in $G$.
                \item[Reference] \cite{Babic1993}
            \end{description}
        \subsubsection{Enumeration of the $k$ shortest paths in a graph}
            \begin{description}
                \item[Input] A graph $G = (V, E)$ and an integer $k$.
                \item[Output] The $k$ smallest shortest paths in $G$.
                \item[Complexity] $O(k|E|)$ total time.
                \item[Reference] \cite{Azevedo1993}
            \end{description}
        \subsubsection{Enumeration of all constrained quickest paths in a network}
            \begin{description}
                \item[Input] Network $N = (V, E)$ and constraints $L$ and $C$.
                \item[Output] All quickest paths in $N$.
                \item[Complexity] $O(k|V|^2|E|)$ total time.
                \item[Comment] $k$ is the number of solutions. A quickest path is a variant of a shortest path.
                \item[Reference] \cite{Gen-Huey1994}
            \end{description}
        \subsubsection{Enumeration of the $k$ shortest paths in a directed graph}
            \begin{description}
                \item[Input] A directed graph $G = (V, E)$ and an integer $k$.
                \item[Output] The $k$ smallest shortest paths in $G$.
                \item[Complexity] $O(k|E|)$ total time
                \item[Reference] \cite{Eppstein1998}
            \end{description}
        \subsubsection{Enumeration of all minimal path conjunctions in a graph}
            \begin{description}
                \item[Input] A directed graph $G = (V, E)$, $s_1, s_2, t_1 \in V$, $T_2 \subseteq V$, and $\mathcal{P} = \{ (s_1, t_1)\} \cup \{(s_2, t): t\in T_2\}$.
                \item[Output] All minimal path conjunctions in $G$.
                \item[Complexity] Polynomial delay.
                \item[Comment] A path conjunction is a edge subset $E' \subseteq E$ such that for all $(s, t) \in \mathcal{P}$, $s$ is connected to $t$ in the graph $G' = (V, E')$.
                \item[Reference] \cite{Boros2004a}
            \end{description}
        \subsubsection{Enumeration of all st-paths in a graph}
            \begin{description}
                \item[Input] A graph $G=(V,E)$ and $s, v \in V$.
                \item[Output] All st-paths in $G$.
                \item[Complexity] $O(|E| + \sum_{\pi \in \mathcal{P}_{st}(G)}|\pi|)$ total time.
                \item[Comment] $\mathcal{P}_{st}(G)$ is the set of all st-paths in $G$.
                \item[Reference] \cite{Ferreira2012}
            \end{description}
        \subsubsection{Enumeration of all $P_3$'s in a graph}
            \begin{description}
                \item[Input] A graph $G$.
                \item[Output] All of all $P_3$'s in $G$.
                \item[Complexity] $O(|E|^{1.5} + p_3(G))$ total time.
                \item[Comment] $P_3$ is a induced path of $G$ with three vertices and $p_3(G)$ is the number of $P_3$ in $G$.
                \item[Reference] \cite{Hoang2013}
            \end{description}
        \subsubsection{Enumeration of all $P_k$'s in a graph}
            \begin{description}
                \item[Input] A graph $G$ and an integer $k \ge 4$.
                \item[Output] All of all $P_k$'s in $G$.
                \item[Complexity] $O(|V|^{k-1} + p_k(G) + k\cdot c_k(G))$ total time.
                \item[Comment] $P_k$ and $C_k$ are a induced path and cycle of $G$ with $k$ vertices, respectively. $p_k(G)$ and $c_k(G)$ are the number of $P_k$ and $C_k$ in $G$, respectively.
                \item[Reference] \cite{Hoang2013}
            \end{description}
        \subsubsection{Enumeration of all $C_k$'s in a graph}
            \begin{description}
                \item[Input] A graph $G$ and an integer $k \ge 4$.
                \item[Output] All of all $C_k$'s in $G$.
                \item[Complexity] $O(|V|^{k-1} + p_k(G) + c_k(G))$ total time.
                \item[Comment] $P_k$ and $C_k$ are a induced path and cycle of $G$ with $k$ vertices, respectively. $p_k(G)$ and $c_k(G)$ are the number of $P_k$ and $C_k$ in $G$, respectively.
                \item[Reference] \cite{Hoang2013}
            \end{description}
        \subsubsection{Enumeration of all chordless $st$-paths in a graph}
            \begin{description}
                \item[Input] A graph $G=(V,E)$ and two vertices $s, t \in V$.
                \item[Output] All chordless $st$-paths in $G$.
                \item[Complexity] $\tilde{O}(|E| +|V|\cdot P )$ total time.
                \item[Comment] $P$ is the number of chordless $st$-paths in $G$.
                \item[Reference] \cite{Ferreira2014}
            \end{description}
        \subsubsection{Enumeration of all chordless st-paths in a graph}
            \begin{description}
                \item[Input] A graph $G=(V,E)$ and $s, t \in V$.
                \item[Output] All chordless st-paths (from $s$ to $t$) in $G$.
                \item[Complexity] $O(|V|+|E|)$ time per chordless st-path.
                \item[Reference] \cite{Unoc}
            \end{description}
     \subsection{Permutation graph}
        \subsubsection{Enumeration of all connected bipartite permutation graphs with $n$ vertices}
            \begin{description}
                \item[Input] A graph size $n$.
                \item[Output] All connected bipartite permutation graphs.
                \item[Complexity] $O(1)$ time per graph with $O(n)$ space.
                \item[Reference] \cite{Saitoh2010b}
            \end{description}
     \subsection{Pitch}
        \subsubsection{Enumeration of all stories in a graph}
            \begin{description}
                \item[Input] A directed graph $G = (V, E, S, T)$ that has the set of source vertices $S$ and the set of target vertices of $T$.
                \item[Output] All stories in $G$.
                \item[Comment] A \textit{pitch} $P$ of $G$ is a set of arcs $E' \subseteq E$, such that the subgraph $G' =(V', E')$ of $G$, where $V' \subseteq V$ is the set of vertices of G having at least one out-going or in-coming arc in $E'$, is acyclic and for each vertex $w \in V' \setminus S$, $w$ is not a source in $G'$, and for each vertex $w \in V' \setminus T$ ,$w$ is not a target in $G'$. $P$ is a \textit{story} if $P$ is maximal.
                \item[Reference] \cite{Acuna2012}
            \end{description}
        \subsubsection{Enumeration of all pitches}
            \begin{description}
                \item[Input] A directed graph $G = (V, E, S, T)$ that has the set of source vertices $S$ and the set of target vertices of $T$.
                \item[Output] All pitches in $G$.
                \item[Complexity] $O(|V| + |E|)$ delay with $O(|V| + |E|)$ space.
                \item[Comment] A pitch $P$ of $G$ is a set of arcs $E' \subseteq E$, such that the subgraph $G' =(V', E')$ of $G$, where $V' \subseteq V$ is the set of vertices of G having at least one out-going or in-coming arc in $E'$, is acyclic and for each vertex $w \in V' \setminus S$, $w$ is not a source in $G'$, and for each vertex $w \in V' \setminus T$ ,$w$ is not a target in $G'$. Enumeration of all stories (maximal pitches) is still open.
                \item[Reference] \cite{Borassi2013}
            \end{description}
     \subsection{Planar graph}
        \subsubsection{Enumeration of all maximal planar graphs with $n$ vertices}
            \begin{description}
                \item[Input] An integer $n$.
                \item[Output] All maximal planar graph with $n$ vertices.
                \item[Complexity] $O(n^3)$ time per graph with $O(n)$ space.
                \item[Comment] A planar graph with $n$ vertices is \textit{maximal} if it has exactly $3n -6$ edges.
                \item[Reference] \cite{Nakano2004a}
            \end{description}
        \subsubsection{Enumeration of all based floorplans with at most $n$ faces}
            \begin{description}
                \item[Input] An integer $n$.
                \item[Output] All based floorplans with at most $n$ faces.
                \item[Complexity] $O(1)$ time per solution with $O(n)$ space.
                \item[Comment] A planar graph is called a \textit{floorplan} if every face is a rectangle. A \textit{based floorplan} is a floorplan with one designated base line segment on the outer face.
                \item[Reference] \cite{Nakano2001a}
            \end{description}
        \subsubsection{Enumeration of all based floorplans with exactly $n$ faces}
            \begin{description}
                \item[Input] An integer $n$.
                \item[Output] All based floorplans with exactly $n$ faces.
                \item[Complexity] $O(1)$ time per solution with $O(n)$ space.
                \item[Comment] A planar graph is called a \textit{floorplan} if every face is a rectangle. A \textit{based floorplan} is a floorplan with one designated base line segment on the outer face.
                \item[Reference] \cite{Nakano2001a}
            \end{description}
        \subsubsection{Enumeration of all floorplans with exactly $n$ faces}
            \begin{description}
                \item[Input] An integer $n$.
                \item[Output] All floorplans with exactly $n$ faces.
                \item[Complexity] $O(1)$ time per solution with $O(n)$ space.
                \item[Comment] A planar graph is called a \textit{floorplan} if every face is a rectangle.
                \item[Reference] \cite{Nakano2001a}
            \end{description}
        \subsubsection{Enumeration of all internally triconnected planar graphs}
            \begin{description}
                \item[Input] Integers $n$ and $g$.
                \item[Output] All internally triconnected planar graphs with exactly $n$ vertices such that $\kappa(G) = 2$ and the size of each inner face is at most $g$.
                \item[Complexity] $O(n^3)$ time per solution on average with $O(n)$ space.
                \item[Reference] \cite{Zhuang2010a}
            \end{description}
     \subsection{Plane graph}
        \subsubsection{Enumeration of all plane straight-line graphs on a given point set in the plane}
            \begin{description}
                \item[Input] A point set $P$ in the plane.
                \item[Output] All plane straight-line graphs on $P$.
                \item[Complexity] $O(|P|\log|P|)$ time per solution.
                \item[Comment] Use gray code.
                \item[Reference] \cite{Aichholzer2007}
            \end{description}
        \subsubsection{Enumeration of all plane and connected straight-line graphs on a given point set in the plane}
            \begin{description}
                \item[Input] A point set $P$ in the plane.
                \item[Output] All plane and connected straight-line graphs on $P$.
                \item[Complexity] $O(|P|\log|P|)$ time per solution.
                \item[Comment] Use gray code.
                \item[Reference] \cite{Aichholzer2007}
            \end{description}
        \subsubsection{Enumeration of all plane spanning trees on a given point set in the plane}
            \begin{description}
                \item[Input] A point set $P$ in the plane.
                \item[Output] All plane spanning trees on $P$.
                \item[Complexity] $O(|P|\log|P|)$ time per solution.
                \item[Comment] Use gray code.
                \item[Reference] \cite{Aichholzer2007}
            \end{description}
        \subsubsection{Enumeration of all plane graphs}
            \begin{description}
                \item[Input] $m$: the maximum number of edges.
                \item[Output] All connected rooted plane graphs with at most $m$ edges.
                \item[Complexity] amortized $O(1)$ time per graph with $O(m)$ space.
                \item[Comment] This algorithm does not outputs the entire graph but the difference from previous one.
                \item[Reference] \cite{Yamanaka2009}
            \end{description}
        \subsubsection{Enumeration of all plane graphs}
            \begin{description}
                \item[Input] $m$: the maximum number of edges.
                \item[Output] All connected non-rooted plane graphs with at most $m$ edges.
                \item[Complexity] $O(m^3)$ time per graph with $O(m)$ space.
                \item[Comment] This algorithm does not outputs the entire graph but the difference from previous one.
                \item[Reference] \cite{Yamanaka2009}
            \end{description}
        \subsubsection{Enumeration of all plane graphs on a given point set in the plane}
            \begin{description}
                \item[Input] A fixed point set $P$.
                \item[Output] All plane graphs on $P$.
                \item[Complexity] $O(N)$ total time.
                \item[Comment] $N$ is the number of solutions.
                \item[Reference] \cite{Katoh2009a}
            \end{description}
        \subsubsection{Enumeration of all non-crossing spanning connected graphs on a given point set in the plane}
            \begin{description}
                \item[Input] A fixed point set $P$.
                \item[Output] All non-crossing spanning connected graphs on $P$.
                \item[Complexity] $O(N)$ total time.
                \item[Comment] $N$ is the number of solutions.
                \item[Reference] \cite{Katoh2009a}
            \end{description}
        \subsubsection{Enumeration of all non-crossing spanning trees on a given point set in the plane}
            \begin{description}
                \item[Input] A fixed point set $P$.
                \item[Output] All non-crossing spanning trees on $P$.
                \item[Complexity] $O(N + |P|tri(P))$ total time.
                \item[Comment] $N$ is the number of solutions and $tri(P)$ is the number of triangulations of $P$.
                \item[Reference] \cite{Katoh2009a}
            \end{description}
        \subsubsection{Enumeration of all non-crossing minimally rigid frameworks on a given point set in the plane}
            \begin{description}
                \item[Input] A fixed point set $P$.
                \item[Output] All non-crossing minimally rigid frameworks on $P$.
                \item[Complexity] $O(|P|^2N)$ total time.
                \item[Comment] $N$ is the number of solutions.
                \item[Reference] \cite{Katoh2009a}
            \end{description}
        \subsubsection{Enumeration of all non-crossing perfect matchings on a given point set in the plane}
            \begin{description}
                \item[Input] A fixed point set $P$.
                \item[Output] All non-crossing perfect matchings on $P$.
                \item[Complexity] $O(|P|^{3/2}tri(P) + |P|^{5/2}N)$ total time.
                \item[Comment] $N$ is the number of solutions and $tri(P)$ is the number of triangulations of $P$.
                \item[Reference] \cite{Katoh2009a}
            \end{description}
        \subsubsection{Enumeration of all triconnected rooted plane graphs}
            \begin{description}
                \item[Input] Integers $n$ and $g$.
                \item[Output] All triconnected rooted plane graphs with $n$ vertices, whose each inner face has the length at most $g$.
                \item[Complexity] $O(1)$ delay with $O(n)$ space after $O(n)$ time preprocessing.
                \item[Reference] \cite{Zhuang2010}
            \end{description}
        \subsubsection{Enumeration of all triconnected rooted plane graphs}
            \begin{description}
                \item[Input] An integer $n$.
                \item[Output] All triconnected rooted plane graphs with $n$ vertices.
                \item[Complexity] $O(n^3)$ delay with $O(n)$ space after $O(n)$ time preprocessing.
                \item[Reference] \cite{Zhuang2010}
            \end{description}
        \subsubsection{Enumeration of all biconnected rooted plane graphs}
            \begin{description}
                \item[Input] Integers $n$ and $g$.
                \item[Output] All biconnected rooted plane graphs with exactly $n$ vertices such that each inner face is of length at most $g$.
                \item[Complexity] $O(1)$ delay with $O(n)$ space, after an $O(n)$ time preprocessing.
                \item[Reference] \cite{Zhuang2010c}
            \end{description}
        \subsubsection{Enumeration of all biconnected plane graphs}
            \begin{description}
                \item[Input] Integers $n$ and $g$.
                \item[Output] All biconnected plane graphs with at most $n$ vertices such that each inner face is of length at most $g$.
                \item[Complexity] $O(n^3)$ time per solution on average with $O(n)$ space.
                \item[Reference] \cite{Zhuang2010c}
            \end{description}
        \subsubsection{Enumeration of all biconnected rooted plane graphs}
            \begin{description}
                \item[Input] Integers $n$ and $g$.
                \item[Output] All biconnected rooted plane graphs with at most $n$ vertices such that each inner face is of length at most $g$.
                \item[Complexity] $O(1)$ delay with $O(n)$ space.
                \item[Reference] \cite{Zhuang2010c}
            \end{description}
        \subsubsection{Enumeration of all rooted plane graphs in $\mathcal{G}_{\text{int}}(n, g) - \mathcal{G}_3(n, g)$}
            \begin{description}
                \item[Input] Integers $n$ and $g$.
                \item[Output] All rooted plane graphs in $\mathcal{G}_{\text{int}}(n, g) - \mathcal{G}_3(n, g)$
                \item[Complexity] $O(1)$ delay with $O(n)$ space and time preprocessing.
                \item[Reference] \cite{Zhuang2010a}
            \end{description}
     \subsection{Polytope}
        \subsubsection{Enumeration of all 3-polytopes of a graph}
            \begin{description}
                \item[Input] A graph $G = (V, E)$.
                \item[Output] All 3-polytopes of $G$.
                \item[Reference] \cite{Deza1993}
            \end{description}
     \subsection{Quadrangle}
        \subsubsection{Enumeration of all quadrangles in a graph}
            \begin{description}
                \item[Input] A connected graph $G = (V, E)$.
                \item[Output] All quadrangles in $G$.
                \item[Complexity] $O(\alpha(G)|E|)$ total time and $O(|E|)$ space.
                \item[Comment] $\alpha(G)$ is the minimum number of edge-disjoint spanning forests into which $G$ can be decomposed.
                \item[Reference] \cite{Chiba1985}
            \end{description}
     \subsection{Quadrangulation}
        \subsubsection{Enumeration of all based biconnected plane quadrangulations with at most $f$ faces}
            \begin{description}
                \item[Input] An integer $f$.
                \item[Output] All based biconnected plane quadrangulations with at most $f$ faces.
                \item[Complexity] $O(1)$ time per quadrangulation and $O(f)$ space.
                \item[Comment] A \textit{plane quadrangulation} is a plane graph such that each inner face has exactly four edges on its contour. A \textit{based plane quadrangulation} is a plane quadrangulation with one designated edge on the outer face.
                \item[Reference] \cite{Li2003}
            \end{description}
     \subsection{Regular graph}
        \subsubsection{Enumeration of all cubic graphs with less than or equal to $n$ vertices}
            \begin{description}
                \item[Input] An integer $n$. 
                \item[Output] All cubic graphs with less than or equal to $n$ vertices. 
                \item[Reference] \cite{Brinkmann2011}
            \end{description}
     \subsection{Series-parallel}
        \subsubsection{Enumeration of all series-parallel graphs with at most $m$ edges}
            \begin{description}
                \item[Input] An integer $m$.
                \item[Output] All series-parallel graphs with at most $m$ edges.
                \item[Complexity] $O(m)$ time per graph.
                \item[Reference] \cite{Kawano2005a}
            \end{description}
     \subsection{Spanning subgraph}
        \subsubsection{Enumeration of all minimal $k$-vertex connected spanning subgraphs in a $k$-connected graph.}
            \begin{description}
                \item[Input] A $k$-connected graph $G$.
                \item[Output] All minimal $k$-vertex connected spanning subgraphs in $G$.
                \item[Complexity] $O(K^3|E|^3|n| + K^2|E|^5|V|^4 + K|V|^k|E|^2)$ total time.
                \item[Comment] $K$ is the number of solutions.  A graph $G$ is $k$-connected if a subgraph of $G$ obtained by removing at most $k-1$ vertices is still connected.
                \item[Reference] \cite{Boros2007}
            \end{description}
     \subsection{Spanning tree}
        \subsubsection{Enumeration of all spanning trees in a graph}
            \begin{description}
                \item[Input] A graph $G$. 
                \item[Output] All spanning trees in  $G$. 
                \item[Complexity]  
                \item[Reference] \cite{Hakimi1961}
            \end{description}
        \subsubsection{Enumeration of all spanning trees in a graph}
            \begin{description}
                \item[Input] A graph $G=(V,E)$.
                \item[Output] All spanning trees of $G$.
                \item[Comment] In this paper, 'trees' indicate 'spanning trees'. 
                \item[Reference] \cite{Minty1965}
            \end{description}
        \subsubsection{Enumeration of all spanning trees of a graph}
            \begin{description}
                \item[Input] A graph $G$. 
                \item[Output] All spanning trees of $G$. 
                \item[Reference] \cite{Mayeda1965a}
            \end{description}
        \subsubsection{Enumeration of all spanning trees in a graph}
            \begin{description}
                \item[Input] A graph $G = (V, E)$.
                \item[Output] All spanning trees in $G$.
                \item[Complexity] $O(|V||E|^2)$ time per spanning tree with $O(|V||E|)$ space.
                \item[Reference] \cite{Read1975}
            \end{description}
        \subsubsection{Enumeration of the $k$ smallest weight spanning trees in a graph}
            \begin{description}
                \item[Input] A graph $G = (V, E)$ and an integer $k$.
                \item[Output] The $k$ smallest weight spanning trees in $G$.
                \item[Complexity] $O(k|E|\alpha (|E|,|V|) + |E|\log |E|)$ total time and $O(k + |E|)$ space, $\alpha(\cdot)$ is Tarjan's inverse of Ackermann's function.
                \item[Reference] \cite{Gabow1977}
            \end{description}
        \subsubsection{Enumeration of all spanning trees in a graph}
            \begin{description}
                \item[Input] A graph $G = (V, E)$ and an integer $k$.
                \item[Output] The all spanning trees in $G$ in order.
                \item[Complexity] $O(N|V|)$ total time and $O(N + |E|)$ space, $N$ is the number of spanning trees in $G$.
                \item[Reference] \cite{Gabow1977}
            \end{description}
        \subsubsection{Enumeration of all spanning trees in an undirected graph}
            \begin{description}
                \item[Input] An undirected graph $G = (V, E)$.
                \item[Output] All spanning trees in $G$.
                \item[Complexity] $O(|V| + |E| + |V|N)$ total time and $O(|V| + |E|)$ space.
                \item[Comment] $N$ is the number of spanning trees in $G$.
                \item[Reference] \cite{Gabow1978}
            \end{description}
        \subsubsection{Enumeration of all spanning trees in a directed graph}
            \begin{description}
                \item[Input] A directed graph $G = (V, E)$.
                \item[Output] All spanning trees in $G$.
                \item[Complexity] $O(|V| + |E| + |E|N)$ total time and $O(|V| + |E|)$ space.
                \item[Comment] $N$ is the number of spanning trees in $G$.
                \item[Reference] \cite{Gabow1978}
            \end{description}
        \subsubsection{Enumeration of the $k$ smallest weight spanning trees in a graph}
            \begin{description}
                \item[Input] A graph $G = (V, E)$ and an integer $k$.
                \item[Output] The $k$ smallest weight spanning trees in $G$.
                \item[Complexity] $O(k|E| + \min (|V|^2 ,|E|\log \log |V|))$ total time and $O(k + |E|)$ space
                \item[Reference] \cite{Katoh1981}
            \end{description}
        \subsubsection{Enumeration of all spanning trees in an undirected graph}
            \begin{description}
                \item[Input] An undirected graph $G$.
                \item[Output] All spanning trees in $G$.
                \item[Comment] They analized Char's enumeration algorithm.
                \item[Reference] \cite{Jayakumar1984}
            \end{description}
        \subsubsection{Enumeration of the $k$ smallest weight spanning trees in a graph in increasing order}
            \begin{description}
                \item[Input] A graph $G = (V, E)$ and an integer $k$.
                \item[Output] The $k$ smallest weight spanning trees in $G$.
                \item[Complexity] $O(|E|\log\log_{(2+|E|/|V|)} n + k^2\sqrt{|E|})$ total time and $O(|E| + k\sqrt{|E|})$ space.
                \item[Reference] \cite{Frederickson1985}
            \end{description}
        \subsubsection{Enumeration of the $k$ smallest weight spanning trees in a planar graph in increasing order}
            \begin{description}
                \item[Input] A planar graph $G = (V, E)$ and an integer $k$.
                \item[Output] The $k$ smallest weight spanning trees in $G$.
                \item[Complexity] $O(|V| + k^2(\log |V|)^2)$ total time and $O(|V| + k(\log |V|)^2)$ space.
                \item[Reference] \cite{Frederickson1985}
            \end{description}
        \subsubsection{Enumeration of all undirected minimum spanning trees in an undirected graph}
            \begin{description}
                \item[Input] An undirected graph $G = (V, E)$.
                \item[Output] All undirected minimum spanning trees in $G$.
                \item[Complexity] $O(|E|\log \beta(|E|, |V|))$ total time.
                \item[Comment] $\beta(|E|, |V|) = \min\{i | \log^{(i)}|V| \le |E|/|N|\}$.
                \item[Reference] \cite{Gabow1986a}
            \end{description}
        \subsubsection{Enumeration of all directed minimum spanning trees in an directed graph}
            \begin{description}
                \item[Input] A directed graph $G = (V, E)$.
                \item[Output] All directed minimum spanning trees in $G$.
                \item[Complexity] $O(|V|\log \beta(|E|, |V|))$ total time.
                \item[Comment] $\beta(|E|, |V|) = \min\{i | \log^{(i)}|V| \le |E|/|N|\}$.
                \item[Reference] \cite{Gabow1986a}
            \end{description}
        \subsubsection{Enumeration of all spanning tree in a graph}
            \begin{description}
                \item[Input] A graph $G = (V, E)$.
                \item[Output] All spanning tree of $G$.
                \item[Complexity] $O(|V|+|E|+N)$ total time and $O(|V||E|)$ space.
                \item[Comment] $N$ is the number of spanning trees in $G$.
                \item[Reference] \cite{Kapoor1991}
            \end{description}
        \subsubsection{Enumeration of all spanning tree in an weighted graph}
            \begin{description}
                \item[Input] An weighted graph $G = (V, E)$.
                \item[Output] All spanning tree of $G$ in increasing order of weight.
                \item[Complexity] $O(N\log |V|+|V||E|)$ total time and $O(N + |V|^2|E|)$ space.
                \item[Comment] $N$ is the number of spanning trees in $G$.
                \item[Reference] \cite{Kapoor1991}
            \end{description}
        \subsubsection{Enumeration of all spanning tree in a directed graph}
            \begin{description}
                \item[Input] A directed graph $G = (V, E)$.
                \item[Output] All spanning tree of $G$.
                \item[Complexity] $O(N|V|+|V|^3)$ total time and $O(|V|^2)$ space.
                \item[Comment] $N$ is the number of spanning trees in $G$.
                \item[Reference] \cite{Kapoor1991}
            \end{description}
        \subsubsection{Enumeration of the $k$ best spanning trees in a graph}
            \begin{description}
                \item[Input] A graph $G = (V, E)$ and an integer $k$.
                \item[Output] The $k$ best spanning trees of $G$.
                \item[Complexity] $O(m\log\beta(|E|, |V|) + k^2)$ total time.
                \item[Comment] $\beta(|E|, |V|) = \min \{\log^{(i)}|V| \le |E|/|V|\}$.
                \item[Reference] \cite{Eppstein1992}
            \end{description}
        \subsubsection{Enumeration of the $k$ best spanning trees in a planar graph}
            \begin{description}
                \item[Input] A planar graph $G = (V, E)$ and an integer $k$.
                \item[Output] The $k$ best spanning trees of $G$.
                \item[Complexity] $O(n + k^2)$ total time
                \item[Reference] \cite{Eppstein1992}
            \end{description}
        \subsubsection{Generation of the $k$-th minimum spanning tree in a graph}
            \begin{description}
                \item[Input] A graph $G = (V, E)$ and an integer $k$.
                \item[Output] The $k$-th minimum spanning tree of $G$.
                \item[Complexity] $O((|V||E|)^{k-1})$ time.
                \item[Reference] \cite{Mayr1992}
            \end{description}
        \subsubsection{Enumeration of all spanning trees}
            \begin{description}
                \item[Input] A graph $G = (V, E)$.
                \item[Output] All spanning trees in $G$.
                \item[Complexity] $O(N + |V| + |E|)$ total time and $O(|V||E|)$ space.
                \item[Comment] $N$ is the number of spanning trees in $G$.
                \item[Reference] \cite{Shioura1995}
            \end{description}
        \subsubsection{Enumeration of all spanning trees in a directed graph}
            \begin{description}
                \item[Input] A directed graph $G = (V, E)$.
                \item[Output] All spanning trees in $G$.
                \item[Complexity] INCORRECT: $O(N \log |V| + |V|^2\alpha(V, V) + |V||E|)$
                \item[Comment] $\alpha$: the inverse Ackermann's function.  [KR2000] gives this result is wrong.
                \item[Reference] \cite{Hariharan1995}
            \end{description}
        \subsubsection{Enumeration of all spanning trees in a directed graph}
            \begin{description}
                \item[Input] A directed graph $G = (V, E)$.
                \item[Output] All spanning trees in $G$.
                \item[Complexity] $O(|E|+ND(|V|, |E|))$ total time and $O(|E|+DS(|V|, |E|))$ space.
                \item[Comment] $D(|V|, |E|)$ and $DS(|V|, |E|)$ are the time and space complexities of the data structure for updating the minimum spanning tree in an undirected graph with $|V|$ vertices and $|E|$ edges. Here $N$ denotes the number of directed spanning trees in $G$.
                \item[Reference] \cite{Uno1996}
            \end{description}
        \subsubsection{Enumeration of all spanning trees in a graph}
            \begin{description}
                \item[Input] A graph $G=(V,E)$.
                \item[Output] All spanning trees included in $G$.
                \item[Complexity] $O(N+|V|+|E|)$ total time and $O(|V|+|E|)$ space.
                \item[Comment] $N$ = number of spanning trees in $G$.
                \item[Reference] \cite{Shioura1997}
            \end{description}
        \subsubsection{Enumeration of the $k$ smallest weight spanning trees in a graph}
            \begin{description}
                \item[Input] A graph $G = (V, E)$ and an integer $k$.
                \item[Output] The $k$ smallest weight spanning trees in $G$.
                \item[Complexity] $O(m \log \log* n + k \min(n, k)^{1/2})$ total time, or a randomized version taking $O(m + k \min(n, k)^{1/2})$ total time.
                \item[Reference] \cite{Eppstein1997}
            \end{description}
        \subsubsection{Enumeration of all spanning trees in a graph}
            \begin{description}
                \item[Input] A graph $G = (V, E)$.
                \item[Output] All spanning trees in $G$.
                \item[Complexity] $O(|V| + |E| + \tau)$ time and $O(|V| + |E|)$ space (depth first manner) or $O(\tau|V| + |E|)$ space (breadth first manner).
                \item[Comment] By using breadth first manner, the proposed algorithm can be used in a parallel computer.
                \item[Reference] \cite{Matsui1997}
            \end{description}
        \subsubsection{Enumeration of all spanning trees in a graph in non-decreasing order}
            \begin{description}
                \item[Input] A graph $G = (V, E)$.
                \item[Output] All spanning trees in $G$ in non-decreasing order.
                \item[Complexity] $O(|V| + |E| + \tau)$ time and $O(\tau|V| + |E|)$ space.
                \item[Comment] Using breadth first manner.
                \item[Reference] \cite{Matsui1997}
            \end{description}
        \subsubsection{Enumeration of all directed spanning trees in a directed graph}
            \begin{description}
                \item[Input] A directed graph $G = (V, E)$.
                \item[Output] All directed spanning trees in $G$.
                \item[Complexity] $O(|E| \log |V| + |V| + N \log^2 |V|)$ total time and $O(|E| + |V|)$ space.
                \item[Comment] $N$ is the number of directed spanning trees.
                \item[Reference] \cite{Uno1998a}
            \end{description}
        \subsubsection{Enumeration of all spanning trees of an weighted graph in order of increasing cost}
            \begin{description}
                \item[Input] An weighted graph $G = (V, E)$.
                \item[Output] All spanning trees of $G$ in order of increasing cost.
                \item[Complexity] $O(N|E|\log |E| + N^2)$ total time and $O(N|E|)$ space.
                \item[Comment] $N$ is the number of spanning trees of $G$.
                \item[Reference] \cite{Sorensen2005}
            \end{description}
        \subsubsection{Enumeration of all the minimum spanning trees in a graph}
            \begin{description}
                \item[Input] An weighted graph $G = (V, E)$.
                \item[Output] All the minimum spanning trees in $G$.
                \item[Complexity] $O(N|E|\log|V|)$ total time and $O(|E|)$ space.
                \item[Comment] $N$ is the number of the minimum spanning trees in $G$.
                \item[Reference] \cite{Yamada2010}
            \end{description}
     \subsection{Steiner tree}
        \subsubsection{Enumeration of all Steiner $W$-trees in a connected graph}
            \begin{description}
                \item[Input] A connected graph $G = (V, E)$, a vertex set $W \subseteq V$ such that $|W| = k$, for a fixed integer $k$.
                \item[Output] Enumeration of all Steiner $W$-trees in $G$.
                \item[Complexity] $O(|V|^2(|V|+|E|) + |V|^{k-2} + N|V|)$ total time with $O(|V|^{k-2} + |V|^2(|V| + |E|))$ space.
                \item[Comment] A connected subgraph $T$ of $G$ is a Steiner $W$-tree if $W \subseteq V(T)$ and $|E(T)$ is minimum.
                \item[Reference] \cite{Dourado2014}
            \end{description}
     \subsection{Subforest}
        \subsubsection{Enumeration of all $k$-trees in a graph}
            \begin{description}
                \item[Input] A graph $G = (V, E)$. 
                \item[Output] All $k$-trees in $G$. 
                \item[Comment] A $k$-tree is a forest with $k$ connected components. 
                \item[Reference] \cite{Hakimi1964}
            \end{description}
     \subsection{Subgraph}
        \subsubsection{Enumeration of all connected induced subgraph of a graph}
            \begin{description}
                \item[Input] A graph $G = (V, E)$.
                \item[Output] All connected induced subgraph of $G$.
                \item[Complexity] $O(|V||E|N)$ total time and $O(|V| + |E|)$ space.
                \item[Comment] $N$ is the number of solutions.
                \item[Reference] \cite{Avis1996}
            \end{description}
        \subsubsection{Enumeration of all connected common maximal subgraphs in two graphs}
            \begin{description}
                \item[Input] Two graphs $G$ and $G'$.
                \item[Output] All connected common maximal subgraphs in $G$ and $G'$.
                \item[Reference] \cite{Koch2001}
            \end{description}
        \subsubsection{Enumeration of all minimal spanning graph}
            \begin{description}
                \item[Input] A graph $G = (V, E)$, $S \subseteq V$, and requirements $r(u, v)$ for all $(u, v) \in V \times V$.
                \item[Output] All minimal spanning graph $H$ of $G$ satisfying $\lambda^S_H \ge r(u, v) \; \forall (u, v) \in V \times V$.
                \item[Complexity] Incremental polynomial time.
                \item[Comment] The $S$-connectivity $\lambda^S_G(u, v)$ of $(u, v)$ in $G$ is the maximum number of $uv$-paths such that no two of them have an edge or a node in $S \setminus \{u, v\}$ in common. This complexity holds for edge-connectivity.
                \item[Reference] \cite{Nutov2009}
            \end{description}
        \subsubsection{Enumeration of all $k$-outconnected minimal spanning graph}
            \begin{description}
                \item[Input] A graph $G = (V, E)$, a vertex $s \in V$, and an integer $k$.
                \item[Output] All minimal $k$-outconneted from $s$ spanning subgraph of $G$.
                \item[Complexity] Incremental polynomial time.
                \item[Comment] A graph is $k$-outconnected from $s$ if it contains $k$ internally-disjoint $st$-paths for every $t \in V$. This complexity holds for both vertex and edge-connectivity.
                \item[Reference] \cite{Nutov2009}
            \end{description}
        \subsubsection{Enumeration of all $k$-outconnected minimal spanning graph}
            \begin{description}
                \item[Input] A directed graph $G = (V, E)$, a vertex $s \in V$, and an integer $k$.
                \item[Output] All minimal $k$-outconneted from $s$ spanning subgraph of $G$.
                \item[Complexity] Incremental polynomial time.
                \item[Comment] A graph is $k$-outconnected from $s$ if it contains $k$ internally-disjoint $st$-paths for every $t \in V$. This complexity holds for both vertex and edge-connectivity.
                \item[Reference] \cite{Nutov2009}
            \end{description}
        \subsubsection{Enumeration of all $k$-connected minimal spanning graph}
            \begin{description}
                \item[Input] A directed graph $G = (V, E)$ and an integer $k$.
                \item[Output] All minimal $k$-connected spanning subgraph of $G$.
                \item[Complexity] Incremental polynomial time.
                \item[Comment] This complexity holds for both vertex and edge-connectivity.
                \item[Reference] \cite{Nutov2009}
            \end{description}
     \subsection{Subtree}
        \subsubsection{Enumeration of all subtrees in an input tree}
            \begin{description}
                \item[Input] A tree $T=(V, E)$.
                \item[Output] All subtrees included in $T$.
                \item[Complexity] $O(|V|)$ delay and $O(|V|)$ space.
                \item[Reference] \cite{Ruskey1981}
            \end{description}
        \subsubsection{Enumeration of all $k$-noded subtrees in a tree}
            \begin{description}
                \item[Input] A tree $T$ and an integer $k$.
                \item[Output] All $k$-noded subtrees in $T$.
                \item[Reference] \cite{Hikita1983}
            \end{description}
        \subsubsection{Enumeration of all $k$-subtrees in a graph}
            \begin{description}
                \item[Input] A graph $G=(V, E)$ and a positive integer $k$.
                \item[Output] All $k$-subtrees included in $G$.
                \item[Complexity] $O(sk)$ total time, $O(k)$ amortized time per solution, and $O(|E|)$ space.
                \item[Comment] $s$ = number of $k$-subtrees in $G$, a $k$-subtree means a connected, acyclic, and edge induced subgraph with $k$ vertices.
                \item[Reference] \cite{Ferreira2011}
            \end{description}
        \subsubsection{Enumeration of all $k$-subtrees in an input tree}
            \begin{description}
                \item[Input] A tree $T=(V, E)$ and an integer $k$.
                \item[Output] All $k$-subtrees included in $T$.
                \item[Complexity] $O(1)$ delay and $O(|V|)$ space after $O(|V|)$ time preprocessing.
                \item[Comment] A $k$-subtree is a connected, acyclic, and edge induced subgraph with $k$ vertices.
                \item[Reference] \cite{Wasa2012b}
            \end{description}
        \subsubsection{Enumeration of all $k$-cardinarity subtrees of a tree with $w$ vertices}
            \begin{description}
                \item[Input] An integer $k$ and a tree $T$ with $w$ element, where $k \le w$.
                \item[Output] All subtrees with $k$ vertices of $T$.
                \item[Complexity] $O(Nw^5)$ total time.
                \item[Comment] $N$ is the number of ideals.
                \item[Reference] \cite{Wild2013}
            \end{description}
        \subsubsection{Enumeration of all induced subtrees in a $k$-degenerate graph}
            \begin{description}
                \item[Input] A $k$-degenerate graph $G = (V, E)$.
                \item[Output] All induced subtrees in $G$.
                \item[Complexity] $O(k)$ amortized time per solution with $O(|V| + |E|)$ space and preprocessing time.
                \item[Reference] \cite{Wasa2010}
            \end{description}
     \subsection{Tour}
        \subsubsection{Enumeration of $k$ best solutions to the Chinese postman problem solutions}
            \begin{description}
                \item[Input] A graph $G = (V, E)$.
                \item[Output] $K$ best solutions to the Chinese postman problem.
                \item[Complexity] $O(S( n,m ) + K( n + m + \log k + nT( n + m,m ) ) )$ where $S( s,t )$ denotes the time complexity of an algorithm for ordinary Chinese postman problems and $T( s,t )$ denotes the time complexity of a post-optimal algorithm for non-bipartite matching problems defined on a graph with s vertices and t edges.
                \item[Reference] \cite{Saruwatari1993}
            \end{description}
     \subsection{Tree}
        \subsubsection{Enumeration of all binary trees with fixed number leaves in lexicographically}
            \begin{description}
                \item[Input] An integer $n$.  
                \item[Output] All binary trees with $n$ leaves in lexicographically. 
                \item[Complexity] $O(1)$ time per binary tree. 
                \item[Reference] \cite{Ruskey1977}
            \end{description}
        \subsubsection{Enumeration of all $t$-ary trees with fixed number leaves in lexicographically}
            \begin{description}
                \item[Input] An integer $n$.  
                \item[Output] All $t$-ary trees with $n$ leaves in lexicographically. 
                \item[Complexity] $O(t)$ time per binary tree. 
                \item[Reference] \cite{Ruskey1978a}
            \end{description}
        \subsubsection{Enumeration of all $k$-ary trees with $n$ vertices}
            \begin{description}
                \item[Input] Integers $k$ and $n$. 
                \item[Output] All $k$-ary trees with $n$ vertices
                \item[Complexity] $O(1)$ time per solution
                \item[Reference] \cite{Trojanowski1978}
            \end{description}
        \subsubsection{Enumeration of all binary trees with $n$ vertices}
            \begin{description}
                \item[Input] An integer $n$. 
                \item[Output] All binary trees with $n$ vertices. 
                \item[Reference] \cite{Rotem1978a}
            \end{description}
        \subsubsection{Enumeration of all $k$-ary trees with $n$ vertices}
            \begin{description}
                \item[Input] Integers $k$ and $n$. 
                \item[Output] All $k$-ary trees with $n$ vertices
                \item[Reference] \cite{Zaks1979}
            \end{description}
        \subsubsection{Enumeration of all rooted trees with $n$ vertices}
            \begin{description}
                \item[Input] An integer $n$.
                \item[Output] All rooted trees with $n$ vertices.
                \item[Complexity] $O(1)$ amortized time per solution.
                \item[Reference] \cite{Beyer1980}
            \end{description}
        \subsubsection{Enumeration of all binary trees with $n$ vertices}
            \begin{description}
                \item[Input] An integer $n$.
                \item[Output] All binary trees with $n$ vertices.
                \item[Reference] \cite{Proskurowski1980}
            \end{description}
        \subsubsection{Enumeration of all $k$-ary trees in lexicographically}
            \begin{description}
                \item[Input] An integer $n$. 
                \item[Output] All $k$-ary trees with $n$ internal vertices in lexicographically
                \item[Complexity] $O(1- (k-1)^{k-1}/k^k)^{-1}$ time per solution. This limit is $4/3$ for the binary case. 
                \item[Reference] \cite{Zaks1980}
            \end{description}
        \subsubsection{Enumeration of all regular $k$-ary trees with $n$ ndoes}
            \begin{description}
                \item[Input] Integers $k$ and $n$.
                \item[Output] All regular $k$-ary trees with $n$ ndoes.
                \item[Reference] \cite{Zaks1982}
            \end{description}
        \subsubsection{Enumeration of all binary trees with $n$ vertices in the lexicographic ordering}
            \begin{description}
                \item[Input] An integer $n$.
                \item[Output] All binary trees with $n$ vertices in the lexicographic ordering
                \item[Complexity] $O(1)$ time per solution on average.
                \item[Reference] \cite{Zerling1985}
            \end{description}
        \subsubsection{Enumeration of all binary trees with $n$ vertices}
            \begin{description}
                \item[Input] An integer $n$.
                \item[Output] All binary trees with $n$ vertices.
                \item[Complexity] $O(n)$ time per solution.
                \item[Reference] \cite{Proskurowski1985}
            \end{description}
        \subsubsection{Enumeration of all ordered trees with $n$ internal vertices}
            \begin{description}
                \item[Input] An integer $n$.
                \item[Output] All ordered trees with $n$ internal vertices.
                \item[Reference] \cite{Er1985}
            \end{description}
        \subsubsection{Enumeration of all free trees with $n$ vertices}
            \begin{description}
                \item[Input] An integer $n$.
                \item[Output] All free trees with $n$ vertices.
                \item[Complexity] $O(1)$ time per solution.
                \item[Reference] \cite{Wright1986}
            \end{description}
        \subsubsection{Enumeration of all $t$-ary trees with $n$ vertices}
            \begin{description}
                \item[Input] Integers $t$ and $n$.
                \item[Output] All $t$-ary trees with $n$ vertices.
                \item[Reference] \cite{Er1987}
            \end{description}
        \subsubsection{Enumeration of all binary trees with $n$ vertices}
            \begin{description}
                \item[Input] An integer $n$.
                \item[Output] All binary trees with $n$ vertices.
                \item[Complexity] $O(1)$ time per solution.
                \item[Reference] \cite{Lucas1987}
            \end{description}
        \subsubsection{Enumeration of all trees with $n$ vertices and $m$ leaves}
            \begin{description}
                \item[Input] Integers $n$ and $m$.
                \item[Output] All trees with $n$ vertices and $m$ leaves
                \item[Reference] \cite{Pallo1987}
            \end{description}
        \subsubsection{Enumeration of all $t$-ary trees in A-order}
            \begin{description}
                \item[Input] Integers $t$ and $n$.
                \item[Output] All $t$-ary trees with $n$ vertices in A-order.
                \item[Complexity] $O(1)$ amortized time per solution.
                \item[Reference] \cite{VanBaronaigien1988}
            \end{description}
        \subsubsection{Enumeration of all binary trees with $n$ leaves}
            \begin{description}
                \item[Input] An integer $n$.
                \item[Output] All binary trees with $n$ leaves.
                \item[Complexity] $O(1)$ amortized time per solution.
                \item[Comment] A strong Gray code can be listed in constant average time per solution.
                \item[Reference] \cite{Ruskey1990}
            \end{description}
        \subsubsection{Enumeration of all binary trees with $n$ vertices}
            \begin{description}
                \item[Input] An integer $n$.
                \item[Output] All binary trees with $n$ vertices.
                \item[Complexity] $O(1)$ time per solution.
                \item[Comment] A loopless generation algorithm is an algorithm where the amount of computation to go from one object to the next is $O(1)$.
                \item[Reference] \cite{VanBaronaigien1991}
            \end{description}
        \subsubsection{Enumeration of all $k$-ary trees in natural order}
            \begin{description}
                \item[Input] Two integers $k$ and $n$.
                \item[Output] All $k$-ary trees with $n$ vertices.
                \item[Reference] \cite{Er1992}
            \end{description}
        \subsubsection{Enumeration of all binary trees with $n$ vertices}
            \begin{description}
                \item[Input] An integer $n$.
                \item[Output] All binary trees with $n$ vertices.
                \item[Complexity] $O(1)$ delay.
                \item[Reference] \cite{Lucas1993}
            \end{description}
        \subsubsection{Enumeration of all binary trees with $n$ vertices}
            \begin{description}
                \item[Input] An integer $n$.
                \item[Output] All binary trees with $n$ vertices.
                \item[Reference] \cite{Bapiraju1994}
            \end{description}
        \subsubsection{Enumeration of all $k$-ary tree}
            \begin{description}
                \item[Input] An integer $k$.
                \item[Output] All $k$-ary trees.
                \item[Complexity] $O(1)$ delay.
                \item[Reference] \cite{Korsh1994}
            \end{description}
        \subsubsection{Enumeration of all binary trees}
            \begin{description}
                \item[Output] All binary trees.
                \item[Complexity] $O(1)$ time per tree.
                \item[Reference] \cite{Xiang1997}
            \end{description}
        \subsubsection{Enumeration of all $k$-ary trees with $n$ vertices}
            \begin{description}
                \item[Input] Two integers $k$ and $n$.
                \item[Output] All $k$-ary trees with $n$ vertices.
                \item[Complexity] $O(1)$ delay.
                \item[Comment] Shifts and loopless algorithm.
                \item[Reference] \cite{Korsh1998}
            \end{description}
        \subsubsection{Enumeration of all rooted plane trees with at most $n$ vertices}
            \begin{description}
                \item[Input] An integer $n$.
                \item[Output] All rooted plane trees with at most $n$ vertices.
                \item[Complexity] $O(1)$ time per tree with $O(n)$ space.
                \item[Comment] A \textit{rooted plane tree} is a rooted tree with a left-to-right ordering specified for the children of each vertex.
                \item[Reference] \cite{Nakano2002}
            \end{description}
        \subsubsection{Enumeration of all rooted plane trees with exactly $n$ vertices}
            \begin{description}
                \item[Input] An integer $n$.
                \item[Output] All rooted plane trees with exactly $n$ vertices.
                \item[Complexity] $O(1)$ time per tree with $O(n)$ space.
                \item[Comment] A \textit{rooted plane tree} is a rooted tree with a left-to-right ordering specified for the children of each vertex.
                \item[Reference] \cite{Nakano2002}
            \end{description}
        \subsubsection{Enumeration of all rooted plane trees with at most $n$ vertices and the maximum degree $D$}
            \begin{description}
                \item[Input] Integers $n$ and $D$.
                \item[Output] All rooted plane trees with at most $n$ vertices and the maximum degree $D$.
                \item[Complexity] $O(1)$ time per tree with $O(n)$ space.
                \item[Comment] A \textit{rooted plane tree} is a rooted tree with a left-to-right ordering specified for the children of each vertex.
                \item[Reference] \cite{Nakano2002}
            \end{description}
        \subsubsection{Enumeration of all rooted plane trees with exactly $n$ vertices and exactly $c$ leaves}
            \begin{description}
                \item[Input] An integer $n$.
                \item[Output] All rooted plane trees with exactly $n$ vertices and exactly $c$ leaves .
                \item[Complexity] $O(n-c)$ time per tree with $O(n)$ space.
                \item[Comment] A \textit{rooted plane tree} is a rooted tree with a left-to-right ordering specified for the children of each vertex.
                \item[Reference] \cite{Nakano2002}
            \end{description}
        \subsubsection{Enumeration of all plane trees with exactly $n$ vertices}
            \begin{description}
                \item[Input] An integer $n$.
                \item[Output] All plane trees with exactly $n$ vertices.
                \item[Complexity] $O(n^3)$ time per tree with $O(n)$ space.
                \item[Comment] A \textit{plane tree} is a tree with a left-to-right ordering specified for the children of each vertex.
                \item[Reference] \cite{Nakano2002}
            \end{description}
        \subsubsection{Enumeration of all rooted trees with at most $n$ vertices}
            \begin{description}
                \item[Input] An integer $n$.
                \item[Output] All rooted tree with at most $n$ vertices.
                \item[Complexity] $O(1)$ time per tree and $O(n)$ space.
                \item[Reference] \cite{Nakano2003}
            \end{description}
        \subsubsection{Enumeration of all $n$-trees}
            \begin{description}
                \item[Input] An integer $n$.
                \item[Output] All $n$-trees.
                \item[Complexity] $O(n^4N)$ total time.
                \item[Comment] Reverse search. $N$ is the number of solutions.
                \item[Reference] \cite{Aringhieri2003}
            \end{description}
        \subsubsection{Enumeration of all trees with $n$ vertices and $d$ diameter}
            \begin{description}
                \item[Input] Integers $n$ and $d$.
                \item[Output] All trees with $n$ vertices and $d$ diameter.
                \item[Complexity] $O(1)$ time per tree with $O(n)$ space.
                \item[Comment] By using the algorithm for each $d = 2, \dots, n-1$, all trees can be enumerated.
                \item[Reference] \cite{Nakano2004}
            \end{description}
        \subsubsection{Enumeration of all $c$-tree with at most $v$ vertices and diameter $d$}
            \begin{description}
                \item[Input] Integers $n$ and $d$
                \item[Output] All $c$-tree with at most $v$ vertices and diameter $d$.
                \item[Complexity] $O(1)$ time per tree.
                \item[Comment] A tree is a $c$-tree if each vertex has a color $c \in \{c_1, \dots, c_m\}$.
                \item[Reference] \cite{Nakano2005a}
            \end{description}
        \subsubsection{Enumeration of all nonisomorphic rooted plane trees with $n$ vertices}
            \begin{description}
                \item[Input] An integer $n$.
                \item[Output] All nonisomorphic rooted plane trees with $n$ vertices.
                \item[Complexity] Constant amortized time per solution.
                \item[Reference] \cite{Sawada2006}
            \end{description}
        \subsubsection{Enumeration of all nonisomorphic free plane trees with $n$ vertices}
            \begin{description}
                \item[Input] An integer $n$.
                \item[Output] All nonisomorphic free plane trees with $n$ vertices.
                \item[Complexity] Constant amortized time per solution.
                \item[Reference] \cite{Sawada2006}
            \end{description}
        \subsubsection{Enumeration of all multitrees satisfying given constraints}
            \begin{description}
                \item[Input] A set $\Sigma$ of labels, a function $val: \Sigma \to \mathbb{Z}^+$, and a feature vector $g$ of level $K$.
                \item[Output] All $(\Sigma, val)$-labeled multitrees $T$ such that $f_K(T) = g$ and $deg(v;T) = val(\ell(v))$ for all vertices $v \in T$.
                \item[Comment] This algorithm is for chemical graphs.
                \item[Reference] \cite{Ishida2008}
            \end{description}
        \subsubsection{Enumeration of all ordered trees with $n$ vertices and $k$ leaves}
            \begin{description}
                \item[Input] Integers $n$ and $k$.
                \item[Output] All ordered trees with $n$ vertices and $k$ leaves.
                \item[Complexity] $O(1)$ delay and $O(n)$ space.
                \item[Reference] \cite{Yamanaka2009a}
            \end{description}
        \subsubsection{Enumeration of all trees with specified degree sequence}
            \begin{description}
                \item[Input] A degree sequence $D$.
                \item[Output] All trees with $D$.
                \item[Complexity] $O(1)$ time per tree.
                \item[Reference] \cite{Nakano2009}
            \end{description}
     \subsection{Triangle}
        \subsubsection{Enumeration of all minimal triangle graphs with a fixed number vertices}
            \begin{description}
                \item[Input] An integer $n$.  
                \item[Output] All minimal triangle graphs with $n$ vertices. 
                \item[Complexity]  
                \item[Reference] \cite{Bowen1967b}
            \end{description}
        \subsubsection{Enumeration of all triangles in a graph}
            \begin{description}
                \item[Input] A graph $G = (V, E)$.
                \item[Output] All triangles in $G$.
                \item[Complexity] $O(\alpha(G)|E|)$ total time and linear space.
                \item[Comment] $\alpha(G)$ is the minimum number of edge-disjoint spanning forests into which $G$ can be decomposed.  If $G$ is planar, then the time complexity becomes $O(|V|)$.
                \item[Reference] \cite{Chiba1985}
            \end{description}
     \subsection{Triangulation}
        \subsubsection{Enumeration of all triangulations of 2-sphere}
            \begin{description}
                \item[Input] A 2-sphere $G$.  
                \item[Output] All triangulations of $G$. 
                \item[Complexity]  
                \item[Reference] \cite{Bowen1967a}
            \end{description}
        \subsubsection{Enumeration of all $r$-rooted $2$-connected triangulations of a planar graph}
            \begin{description}
                \item[Input] A planar graph $G = (V, E)$ and an integer $r$.
                \item[Output] All $r$-rooted $2$-connected triangulations of $G$.
                \item[Complexity] $O(|V|^2N)$ total time and $O(|V|)$ space.
                \item[Comment] $N$ is the number of solutions.
                \item[Reference] \cite{Avis1996b}
            \end{description}
        \subsubsection{Enumeration of all $r$-rooted $3$-connected triangulations of a planar graph}
            \begin{description}
                \item[Input] A planar graph $G = (V, E)$ and an integer $r$.
                \item[Output] All $r$-rooted $3$-connected triangulations of $G$.
                \item[Complexity] $O(|V|^2N)$ total time and $O(|V|)$ space.
                \item[Comment] $N$ is the number of solutions.
                \item[Reference] \cite{Avis1996b}
            \end{description}
        \subsubsection{Enumeration of all based plane triangulations with $n$ vertices}
            \begin{description}
                \item[Input] An integer $n$.
                \item[Output] All based plane triangulations.
                \item[Complexity] $O(1)$ time per based plane triangulation with $O(n)$ space.
                \item[Comment] A \textit{based plane triangulation} is a plane triangulation with one designated edge on the outer face. The algorithm does not output entire solution but output the difference from the previous solution.
                \item[Reference] \cite{Nakano2004a}
            \end{description}
        \subsubsection{Enumeration of all biconnected based plane triangulations with $n$ vertices and $r$ vertices on the outer face}
            \begin{description}
                \item[Input] Integers $n$ and $r$.
                \item[Output] All biconnected based plane triangulations with $n$ vertices and $r$ vertices on the outer face.
                \item[Complexity] $O(1)$ time per based plane triangulation with $O(n)$ space.
                \item[Comment] A \textit{based plane triangulation} is a plane triangulation with one designated edge on the outer face. The algorithm does not output entire solution but output the difference from the previous solution.
                \item[Reference] \cite{Nakano2004a}
            \end{description}
        \subsubsection{Enumeration of all biconnected plane triangulations with $n$ vertices and $r$ vertices on the outer face}
            \begin{description}
                \item[Input] Integers $n$ and $r$.
                \item[Output] All biconnected based plane triangulations with $n$ vertices and $r$ vertices on the outer face.
                \item[Complexity] $O(r^2n)$ time per based plane triangulation with $O(n)$ space.
                \item[Reference] \cite{Nakano2004a}
            \end{description}
        \subsubsection{Enumeration of all rooted triconnected plane triangulations with at most $n$ vertices}
            \begin{description}
                \item[Input] An integer $n$.
                \item[Output] All triconnected rooted plane triangulations with at most $n$ vertices.
                \item[Complexity] $O(1)$ time per tree and $O(n)$ space.
                \item[Comment] A \textit{rooted} plane triangulation is a plane triangulation with one designated vertex on the outer face.
                \item[Reference] \cite{Nakano2001}
            \end{description}
        \subsubsection{Enumeration of all rooted triconnected plane triangulations with exactly $n$ vertices and $r$ leaves}
            \begin{description}
                \item[Input] An integer $n$.
                \item[Output] All triconnected rooted plane triangulations with exactly $n$ vertices and $r$ leaves.
                \item[Complexity] $O(r)$ time per tree and $O(n)$ space.
                \item[Comment] A \textit{rooted} plane triangulation is a plane triangulation with one designated vertex on the outer face.
                \item[Reference] \cite{Nakano2001}
            \end{description}
        \subsubsection{Enumeration of all triconnected plane triangulations with exactly $n$ vertices and $r$ leaves}
            \begin{description}
                \item[Input] An integer $n$.
                \item[Output] All triconnected plane triangulations with exactly $n$ vertices and $r$ leaves.
                \item[Complexity] $O(r^n)$ time per triangulation and $O(n)$ space.
                \item[Reference] \cite{Nakano2001}
            \end{description}
        \subsubsection{Enumeration of all triangulations}
            \begin{description}
                \item[Input] A set $S$ of $n$ points in the general position in the plane.
                \item[Output] all the triangulations whose vertex set is $S$ and edge set includes the convex hull of $S$.
                \item[Complexity] $O(\log\log n)$ time per triangulation and linear space.
                \item[Comment] Whether there is the algorithm that outputs all triangulations in constant time delay?
                \item[Reference] \cite{Bespamyatnikh2002}
            \end{description}
        \subsubsection{Enumeration of all biconnected plane triangulations with $n$ vertices and $r$ vertices on the outer faces}
            \begin{description}
                \item[Input] Integers $n$ and $r$.
                \item[Output] All biconnected plane triangulations with $n$ vertices and $r$ vertices on the outer faces.
                \item[Complexity] $O(rn)$ time per triangulation and $O(n)$ space.
                \item[Reference] \cite{Nakano2004b}
            \end{description}
        \subsubsection{Enumeration of all triangulations of a triconnected plane graph of $n$ vertices}
            \begin{description}
                \item[Input] A triconnected planar graph $G$ with $n$ vertices.
                \item[Output] All triangulations of $G$.
                \item[Complexity] $O(1)$ time per triangulation and $O(n)$ space.
                \item[Reference] \cite{Parvez2009}
            \end{description}
     \subsection{Vertex cover}
        \subsubsection{Enumeration of all minimal vertex covers in a graph}
            \begin{description}
                \item[Input] A graph $G = (V, E)$.
                \item[Output] All minimal vertex covers of size up to $k$ in $G$.
                \item[Complexity] $O^*(1.6181^k)$ total time.
                \item[Comment] This algorithm also lists some non-minimal vertex covers. This algorithm uses compact representation technique.
                \item[Reference] \cite{Fernau2006}
            \end{description}
        \subsubsection{Enumeration of all minimal vertex covers of size at most $k$ in a graph}
            \begin{description}
                \item[Input] A graph $G = (V, E)$.
                \item[Output] All minimal vertex covers of size at most $k$ in $G$.
                \item[Complexity] $O(|E| + k^2 2^k)$ total time.
                \item[Reference] \cite{Damaschke2006}
            \end{description}
\section{Hypergraph}
     \subsection{Acyclic subhypergraph}
        \subsubsection{Enumeration of all maximal $\alpha$-acyclic subhypergraphs in a hypergraph}
            \begin{description}
                \item[Input] A hypergraph $H = (V, \mathcal{E})$.
                \item[Output] All maximal $\alpha$-acyclic subhypergraphs in $H$.
                \item[Complexity] $O(|\mathcal{E}|^2(|V|+|\mathcal{E}|))$ delay and $O(|\mathcal{E}|)$space.
                \item[Comment] The name of their algorithm is $\mathtt{GenMAS}$. This algorithm uses the algorithm $\mathtt{FindMAS}$ that outputs a maximal $\alpha$-acyclic subhypergraph.
                \item[Reference] \cite{Daigo2009}
            \end{description}
        \subsubsection{Enumeration of all Berge acyclic subhypergraphs in a hypergraph}
            \begin{description}
                \item[Input] A hypergraph $\mathcal{H}$.
                \item[Output] All Berge acyclic subhypergraphs in $\mathcal{H}$.
                \item[Complexity] $O(rd\tau(m))$ time per subhypergraph.
                \item[Comment] $r$ and $d$ are the rank and the degree of $\mathcal{H}$ and $\tau(m) = O((\log\log m)^2 / \log \log \log (m))$.
                \item[Reference] \cite{Wasa2013}
            \end{description}
     \subsection{Independent set}
        \subsubsection{Enumeration of all maximal independent set of a hypergraph of bounded dimension}
            \begin{description}
                \item[Input] A hypergraph $H$ of bounded dimension.
                \item[Output] All maximal independent set of a hypergraph of bounded dimension. 
                \item[Comment] The proposed algorithm runs in parallel.
                \item[Reference] \cite{Boros2000a}
            \end{description}
        \subsubsection{Enumeration of all maximal independent sets in a $c$-conformal hypergraph}
            \begin{description}
                \item[Input] A $c$-conformal hypergraph $\mathcal{H} \in A(k, r)$, where $c \le$ constant and $k + r \le c$.
                \item[Output] All maximal independent sets in $\mathcal{H}$.
                \item[Complexity] Incremental polynomial time.
                \item[Comment] $A(k, r)$ is the class of hyperedges with $(k, r)$-bounded intersections, i.e. in which the intersection of any $k$ distinct hyperedges has size at most $r$.
                \item[Reference] \cite{Boros2004}
            \end{description}
        \subsubsection{Enumeration of all maximal independent sets in a hypergraph of bounded intersections}
            \begin{description}
                \item[Input] A hypergraph $\mathcal{H} \in A(k, r)$, where $k + r \le$ constant.
                \item[Output] All maximal independent sets in $\mathcal{H}$.
                \item[Complexity] Incremental polynomial time with polynomial space.
                \item[Comment] $A(k, r)$ is the class of hyperedges with $(k, r)$-bounded intersections, i.e. in which the intersection of any $k$ distinct hyperedges has size at most $r$.
                \item[Reference] \cite{Boros2004}
            \end{description}
     \subsection{Transversal}
        \subsubsection{Enumeration of all minimal transversal of a hypergraph}
            \begin{description}
                \item[Input] A hypergraph $\mathcal{H}$.
                \item[Output] All minimal transversal of $\mathcal{H}$.
                \item[Complexity] $(n + N)^{O(\log n)}$ total time with $O(n \log n)$ words.
                \item[Comment] $n = \sum_{X \in \mathcal{H}}|X|$ and $N$ is the number of solutions.
                \item[Reference] \cite{Tamaki2000}
            \end{description}
        \subsubsection{Enumeration of all minimal transversal in a hypergraph}
            \begin{description}
                \item[Input] A hypergraph $\mathcal{H}\in A(k, r)$.
                \item[Output] All minimal transversal in $\mathcal{H}$.
                \item[Complexity] $O(n^{k+r+1}|\mathcal{H}^d|^{r+1})$ total time and $O(N^{r+1})$ total space.
                \item[Comment] $A(k, r)$ is the class of hyperedges with $(k, r)$-bounded intersections, i.e. in which the intersection of any $k$ distinct hyperedges has size at most $r$.  Minimal transversals of hypergraphs in some restricted classes can be enumerating in polynomial delay and space.
                \item[Reference] \cite{Boros2004}
            \end{description}
\section{Matroid}
     \subsection{Basis}
        \subsubsection{Enumeration of all common bases in two matroids}
            \begin{description}
                \item[Input] Two matroids  $M_1 = (E, \beta_1)$ and $M_2 = (E, \beta_2)$.
                \item[Output] All $B \in \beta_1 \cap \beta_2$.
                \item[Complexity] $O(|E| ( |E|^2 + t)\lambda)$ total time and $O(|E|^2)$ space.
                \item[Comment] $\lambda$ is the number of common bases and $t$ is time complexity of one pivot operation.
                \item[Reference] \cite{Fukuda1995}
            \end{description}
        \subsubsection{Enumeration of all basis of a matroid}
            \begin{description}
                \item[Input] A matroid $M$ on the ground set $P$ with rank $m$.
                \item[Output] All basis of $M$.
                \item[Complexity] $O((m|P| + t(Piv))N)$ total time and space complexity independent of $N$.
                \item[Comment] $N$ is the number of solutions. $t(Piv)$ is the time necessary to do one pivot operation.
                \item[Reference] \cite{Avis1996}
            \end{description}
     \subsection{Spanning}
        \subsubsection{Enumeration of all minimal spanning and connected subsets in a matroid}
            \begin{description}
                \item[Input] A matroid $M$.
                \item[Output] All minimal spanning and connected subsets in $M$.
                \item[Complexity] Incremental quasi-polynomial time.
                \item[Comment] $f(x)$ is a quasi-polynomial if $f(x) \in O(2^{polylog (n)})$.
                \item[Reference] \cite{Khachiyan2006a}
            \end{description}
     \subsection{Subset}
        \subsubsection{Enumeration of all maximal subset}
            \begin{description}
                \item[Input] A binary matroid $M$ on ground set $S$ and $B = \{b_1, b_2\} \subseteq S$.
                \item[Output] All maximal subsets $X$ of $A := S \setminus B$ that span neither $b_1$ nor $b_2$.
                \item[Complexity] Incremental polynomial time.
                \item[Reference] \cite{Khachiyan2008a}
            \end{description}
\section{Order}
     \subsection{Ideal}
        \subsubsection{Enumeration of all $k$-cardinarity ideals of a $w$-element poset}
            \begin{description}
                \item[Input] An integer $k$ and a poset $P$ with $w$ element, where $k \le w$.
                \item[Output] All $k$-cardinarity ideals of $P$.
                \item[Complexity] $O(Nw^3)$ total time.
                \item[Comment] $N$ is the number of ideals.
                \item[Reference] \cite{Wild2013}
            \end{description}
\section{Other}
     \subsection{Assignment}
        \subsubsection{Enumeration of all assignment}
            \begin{description}
                \item[Input] An integer $n$ and $n \times n$ cost matrix $C = (c_{ij})$.
                \item[Output] All assignments that minimizes $\sum^{i=n}_{i=1}\sum^{j=n}_{j=1} c_{ij}x_{ij}$ subject to $\sum^{i=n}_{i=1} x_{ij} = 1 \; (j = 1, \dots, n)$, $\sum^{j=n}_{j=1} x_{ij} = 1\; (i = 1, \dots, n)$, and $x_{ij} \ge 0$.
                \item[Reference] \cite{Murty1968}
            \end{description}
     \subsection{Full disjunction}
        \subsubsection{Enumeration of all full disjunction in an acyclic set}
            \begin{description}
                \item[Input] An acyclic set of relation $R$ with $N$ tuples.
                \item[Output] All full disjunctions of $R$.
                \item[Complexity] $O(N)$ delay.
                \item[Reference] \cite{Cohen2006}
            \end{description}
     \subsection{Matrix}
        \subsubsection{Enumeration of all minimal sets of at most $k$ rows the deletion of which leaves a PP matrix}
            \begin{description}
                \item[Input] A binary $n \times m$ matrix $B$ and a positive integer $k$, where $n > 4k$.
                \item[Output] All minimal sets of at most $k$ rows the deletion of which leaves a PP matrix.
                \item[Complexity] $O(3^k nm)$ time.
                \item[Comment] A PP matrix is a perfect phylogeny matrix.
                \item[Reference] \cite{Damaschke2006}
            \end{description}
     \subsection{Round-robin tournament score}
        \subsubsection{Enumeration of all round-robin tournament scores of $n$ players}
            \begin{description}
                \item[Input] An integer $n$. 
                \item[Output] All round-robin tournament scores of $n$ players. 
                \item[Reference] \cite{Narayana1971}
            \end{description}
\section{Permutation}
     \subsection{Arrangements}
        \subsubsection{Enumeration of all arrangements with $n$ marks}
            \begin{description}
                \item[Input] An integer $n$. 
                \item[Output] All arrangements with $n$ marks. 
                \item[Complexity] $O(n!)$ total time. 
                \item[Reference] \cite{Wells1961}
            \end{description}
        \subsubsection{Enumeration of all arrangements with $n$ marks}
            \begin{description}
                \item[Input] An integer $n$. 
                \item[Output] All arrangements with $n$ marks. 
                \item[Complexity] $O(n!)$ total time. 
                \item[Reference] \cite{Johnson1963}
            \end{description}
     \subsection{Ladder lottery}
        \subsubsection{Enumeration of all optimal ladder lotteries}
            \begin{description}
                \item[Input] A permutation.
                \item[Output] All optimal ladder lotteries with satisfying the permutation.
                \item[Complexity] $O(1)$ time per solution on average, $O(n^2)$ space, and $O(n^2)$ time preprocessing.
                \item[Comment] $n$ is the length of the input permutation.  We call a ladder lottery is an optimal when the number of horizontal lines in the ladder lottery is minimum. Ladder lotteries are also known as arrangements of pseudolines.
                \item[Reference] \cite{Yamanaka2010}
            \end{description}
        \subsubsection{Enumeration of all ladder lotteries with $k$ bars}
            \begin{description}
                \item[Input] A permutation $\pi$ and integer $k$.
                \item[Output] All ladder lotteries of $\pi$ with $k$ bars.
                \item[Complexity] $O(1)$ time per ladder lottery.
                \item[Comment] Ladder lotteries are also known as \textit{Amida kuji} in Japan.
                \item[Reference] \cite{Yamanaka2014}
            \end{description}
     \subsection{Set}
        \subsubsection{Enumeration of all permutations of a set of elements}
            \begin{description}
                \item[Input] A set $S$. 
                \item[Output] All permutations of $S$
                \item[Comment] He proposed the general algorithm for some combinatorial problems. 
                \item[Reference] \cite{Ehrlich1973a}
            \end{description}
        \subsubsection{Enumeration of all permutations of a set of elements}
            \begin{description}
                \item[Input] A set $S$. 
                \item[Output] All permutations of $S$
                \item[Reference] \cite{Dershowitz1975}
            \end{description}
\section{SAT}
     \subsection{Boolean CSP}
        \subsubsection{Enumeration of all models of $\phi$ by non-decreasing weight}
            \begin{description}
                \item[Input] A $\Gamma$-formula $\phi$.
                \item[Output] All models of $\phi$ by non-decreasing weight.
                \item[Complexity] If $\Gamma$ is Horn or width-2 affine, there exists a polynomial delay algorithm.
                \item[Comment] Otherwise, such an algorithm does not exist unless $P \neq NP$.
                \item[Reference] \cite{Creignou2011}
            \end{description}
\section{Set}
     \subsection{Bitstring}
        \subsubsection{Enumeration of all bitstrings of length $n$ that contains exactly $k$ 1's}
            \begin{description}
                \item[Input] An even integer $n$ and odd integer $k$.
                \item[Output] All bitstrings of length $n$ that contains exactly $k$ 1's.
                \item[Complexity] $O(1)$ amortized time per solution.
                \item[Reference] \cite{Ruskey1988}
            \end{description}
     \subsection{Ideals}
        \subsubsection{Enumeration of all ideals in a poset}
            \begin{description}
                \item[Input] A poset $\mathcal{P}$.
                \item[Output] All ideals in $\mathcal{p}$.
                \item[Complexity] $O(1)$ delay.
                \item[Reference] \cite{Koda1993}
            \end{description}
     \subsection{Partition}
        \subsubsection{Enumeration of all paritions in natural order}
            \begin{description}
                \item[Input] An integer $n$. 
                \item[Output] All partitions of $n$ in natural order
                \item[Reference] \cite{McKay1970}
            \end{description}
        \subsubsection{Enumeration of all paritions with restriction}
            \begin{description}
                \item[Input] Integers $k$ and $n$. 
                \item[Output] All partitions of $n$ whose the smallest part is greater than or equal to $k$. 
                \item[Reference] \cite{White1970}
            \end{description}
        \subsubsection{Enumeration of all $k$-partitions of $n$}
            \begin{description}
                \item[Input] Integers $k$ and $n$. 
                \item[Output] All $k$-partitions of $n$. 
                \item[Reference] \cite{Narayana1971}
            \end{description}
        \subsubsection{Enumeration of all paritions of an integer}
            \begin{description}
                \item[Input] An integer $n$.  
                \item[Output] All paritions of $n$. 
                \item[Complexity]  
                \item[Reference] \cite{Fenner1980}
            \end{description}
        \subsubsection{Enumeration of all partitions of a set}
            \begin{description}
                \item[Input] A set $S$.
                \item[Output] All partitions of $S$.
                \item[Complexity] $O(1)$ amortized time per solution.
                \item[Reference] \cite{Semba1984}
            \end{description}
        \subsubsection{Enumeration of all partions of $n$ into integers of size at most $k$}
            \begin{description}
                \item[Input] Integers $n$ and $k$.
                \item[Output] All partions of $n$ into integers of size at most $k$.
                \item[Reference] \cite{Savage1989}
            \end{description}
        \subsubsection{Enumeration of all partitions of a set into a fixed number of blocks}
            \begin{description}
                \item[Input] An integer $k$.
                \item[Output] All partitions of a set into $k$ blocks
                \item[Complexity] $O(1)$ amortized time per solution.
                \item[Reference] \cite{Ruskey1993}
            \end{description}
        \subsubsection{Enumeration of all partitions of an integer $n$}
            \begin{description}
                \item[Input] Three integers $n$, $k$, and $\sigma$.
                \item[Output] All partitions $P_\sigma(n, k)$ of $n$ into parts of size at most $k$ in which parts are congruent to $1$ modulo $\sigma$.
                \item[Complexity] $O(N)$ total time. E.g., $P_3(11, 8) = P_3(11, 7) = \left\{ \{7, 4\}, \{7, 1, 1, 1, 1\}, \{4, 4, 1, 1, 1\}, \{4, 1, 1, 1, 1, 1, 1, 1\}, \{1, \dots, 1\}\right\}$.
                \item[Comment] $N$ is the number of partitions.
                \item[Reference] \cite{Rasmussen1995}
            \end{description}
        \subsubsection{Enumeration of all partitions of an integer $n$}
            \begin{description}
                \item[Input] Two integers $n$ and $k$.
                \item[Output] All partitions $D(n, k)$ of $n$ into distinct parts of size at most $k$.
                \item[Complexity] $O(N)$ total time. E.g., $D(10, 5) = \{\{5, 4, 1\}, \{5, 3, 2\}, \{4, 3, 2, 1\}\}$ and $D(11, 4) = \emptyset$.
                \item[Comment] $N$ is the number of partitions.
                \item[Reference] \cite{Rasmussen1995}
            \end{description}
        \subsubsection{Enumeration of all partition of $\{1, \dots, n\}$ into $k$ non-empty subsets}
            \begin{description}
                \item[Input] An integer $k$.
                \item[Output] All partition of $\{1, \dots, n\}$ into $k$ non-empty subsets.
                \item[Complexity] $O(1)$ time per solution.
                \item[Comment] The number of such partitions is known as the Stirling number of the second kind.
                \item[Reference] \cite{Kawano2005b}
            \end{description}
        \subsubsection{Enumeration of all integer partitions in (anti-)lexicographical order}
            \begin{description}
                \item[Input] An integer $n$.
                \item[Output] All integer partitions of $n$.
                \item[Complexity] $O(1)$ time per solution on average.
                \item[Reference] \cite{Stojmenovic2007}
            \end{description}
\section{String}
     \subsection{Binary string}
        \subsubsection{Enumeration of all binary string with fixed number ones}
            \begin{description}
                \item[Input] Two integers $n$ and $k$, where $n \ge k$.  
                \item[Output] All binary string with length $n$ and $k$ ones. 
                \item[Complexity] $O(1)$ time per binary string. 
                \item[Reference] \cite{Bitner1976}
            \end{description}
     \subsection{Bracelet}
        \subsubsection{Enumeration of all $k$-ary bracelets}
            \begin{description}
                \item[Input] $n$: a length of a bracelet, $k$: a number of alphabet size.
                \item[Output] All $k$-ary bracelets.
                \item[Complexity] $O(1)$ amortized per output and $O(n)$ space.
                \item[Comment] A bracelet is the lexicographically smallest element of an equivalence class of $k$-ary strings under string rotation and reversal.
                \item[Reference] \cite{Sawada2001}
            \end{description}
     \subsection{Lyndon word}
        \subsubsection{Enumeration of all $k$-ary Lyndon brackets of length $n$}
            \begin{description}
                \item[Input] Two integers $k$ and $n$.
                \item[Output] All $k$-ary Lyndon brackets of length $n$.
                \item[Complexity] $O(n)$ time per solution.
                \item[Reference] \cite{Sawada2003}
            \end{description}
     \subsection{Necklace}
        \subsubsection{Enumeration of all $k$-color necklaces with $n$ beads}
            \begin{description}
                \item[Input] Integers $k$ and $n$. 
                \item[Output] All $k$-color necklaces with $n$ beads
                \item[Reference] \cite{Fredricksen1978a}
            \end{description}
        \subsubsection{Enumeration of all necklaces of length $n$ with two colors}
            \begin{description}
                \item[Input] An integer $n$.
                \item[Output] All necklaces of length $n$ with two colors.
                \item[Reference] \cite{Fredricksen1986b}
            \end{description}
        \subsubsection{Enumeration of all $k$-ary necklaces}
            \begin{description}
                \item[Input] $n$: a length of a necklace, $k$: a number of alphabet size.
                \item[Output] All $k$-ary necklaces with length $n$.
                \item[Complexity] $O(1)$ amortized per output and $O(n)$ space.
                \item[Comment] A $k$-ary necklace is an equivalence class of k-ary strings under rotation.
                \item[Reference] \cite{Ruskey1992}
            \end{description}
        \subsubsection{Enumeration of all $n$-bit necklaces with fixed density $d$}
            \begin{description}
                \item[Input] Two integer $n$ and $d$.
                \item[Output] All $n$-bit necklaces with fixed density $d$.
                \item[Complexity] $O(nN)$ total time and $O(n)$ space.
                \item[Comment] $N$ is the number of solutions. A density of a $n$-bit necklace $T$ is $d$ if $T$ has $d$ ones.
                \item[Reference] \cite{Wang1996}
            \end{description}
        \subsubsection{Enumeration of all $k$-ary necklaces with fixed density}
            \begin{description}
                \item[Input] $n$: a length of a necklace, $k$: a number of alphabet size, $d$: a number of nonzero characters.
                \item[Output] All $k$-ary necklaces with fixed density $d$.
                \item[Complexity] $O(1)$ amortized per output and $O(n)$ space.
                \item[Comment] The set of $3$-ary necklace with $2$-density and $4$-length is $N_3(4,2) = \{0011, 0012, 0021, 0022, 0101, 0102, 0202\}$.
                \item[Reference] \cite{Ruskey1999}
            \end{description}
        \subsubsection{Enumeration of all strings of some family}
            \begin{description}
                \item[Complexity] Constant amortized time per solution.
                \item[Comment] This algorithm can list not only all necklaces but also all strings in other some family with CAT.
                \item[Reference] \cite{Cattell2000}
            \end{description}
        \subsubsection{Enumeration of all $k$-ary necklaces with fixed content of length $n$}
            \begin{description}
                \item[Input] Two integer $k$ and $n$, and a content.
                \item[Output] All $k$-ary necklaces with fixed content of length $n$.
                \item[Complexity] Constant amortized time per solution.
                \item[Reference] \cite{Sawada2003a}
            \end{description}
     \subsection{Parenthesis}
        \subsubsection{Enumeration of all well-formed parenthesis with length $2n$}
            \begin{description}
                \item[Input] An integer $n$.
                \item[Output] All well-formed parenthesis with length $2n$ in lexicographical ordering.
                \item[Reference] \cite{Er1983}
            \end{description}
        \subsubsection{Enumeration of all well-formed parenthesis strings of size $n$}
            \begin{description}
                \item[Input] An integer $n$.
                \item[Output] All well-formed parenthesis strings of size $n$.
                \item[Complexity] $O(1)$ delay with $O(n)$ space, or $O(n)$ delay with $O(1)$ space.
                \item[Reference] \cite{Walsh1998}
            \end{description}
     \subsection{Substring}
        \subsubsection{Enumeration of all $k$-ary strings of length $n$ that have no substring equal to $f$}
            \begin{description}
                \item[Input] A $k$-ary string $f$ with length $m$, and a positive integer $n$.
                \item[Output] All $k$-ary strings of length $n$ that have no substring equal to $f$.
                \item[Complexity] $O(1)$ time per string.
                \item[Reference] \cite{Ruskey2000a}
            \end{description}
        \subsubsection{Enumeration of all circular $k$-ary strings of length $n$ that have no substring equal to $f$}
            \begin{description}
                \item[Input] A $k$-ary string $f$ with length $m$, and a positive integer $n$.
                \item[Output] All circular $k$-ary strings of length $n$ that have no substring equal to $f$.
                \item[Complexity] $O(1)$ time per string.
                \item[Reference] \cite{Ruskey2000a}
            \end{description}
        \subsubsection{Enumeration of all $k$-ary necklaces of length $n$ that have no substring equal to $f$}
            \begin{description}
                \item[Input] A $k$-ary string $f$ with length $m$, and a positive integer $n$.
                \item[Output] All $k$-ary necklaces of length $n$ that have no substring equal to $f$.
                \item[Complexity] $O(1)$ time per string.
                \item[Comment] $f$ is an aperiodic necklace.
                \item[Reference] \cite{Ruskey2000a}
            \end{description}
\section{Survey}
     \subsection{Enumeration}
        \subsubsection{Enumeration of $k$-best enumeration}
            \begin{description}
                \item[Comment] This is the full version of the Springer Encyclopedia of Algorithms, 2014.
                \item[Reference] \cite{Eppstein2014}
            \end{description}
     \subsection{Graph}
        \subsubsection{Alglrotihms for enumeration of all cycles in a given graph}
            \begin{description}
                \item[Input] A graph $G$. 
                \item[Output] All cycles belonging to $G$. 
                \item[Comment] 26 cycle enumeration algorithms are introduced. 
                \item[Reference] \cite{Mateti1976a}
            \end{description}
        \subsubsection{Enumeration of clique enumeration}
            \begin{description}
                \item[Comment] In Sec.3.
                \item[Reference] \cite{Pardalos1994}
            \end{description}
        \subsubsection{Enumeration of gray code algorithms}
            \begin{description}
                \item[Reference] \cite{Savage1997}
            \end{description}
     \subsection{Logic}
        \subsubsection{}
            \begin{description}
                \item[Comment] The author of this paper investigate the class of databases with constant delay the answers to a query.
                \item[Reference] \cite{Segoufin2013}
            \end{description}